\documentclass[journal,comsoc]{IEEEtran}
\usepackage[T1]{fontenc}

\usepackage{cite}

\ifCLASSINFOpdf
  \usepackage[pdftex]{graphicx}
\else
\fi
\usepackage{amsmath}
\usepackage[cmintegrals]{newtxmath}
\usepackage{algorithm}
\usepackage[noend]{algpseudocode}

\usepackage{cite}
\usepackage{textcomp}
\usepackage{url}
\usepackage{subcaption}
\usepackage{autobreak}
\usepackage{tabularx}
\usepackage{booktabs}
\usepackage{tabularray}
\usepackage{listings}
\usepackage{multirow}
\usepackage{autobreak}
\usepackage{balance}

\usepackage{mathtools}
\let\oldsum\sum
\renewcommand{\sum}{\futurelet\next\sumCheck}
\def\sumCheck{\ifx\next_\expandafter\sumSmash\else\expandafter\oldsum\fi}
\def\sumSmash_#1{\smashoperator{\oldsum_{#1}}}

\lstdefinestyle{bash}{
  language=bash,
  basicstyle=\ttfamily\small,
  keywordstyle=\color{blue!70},
  commentstyle=\color{gray},
  stringstyle=\color{olive},
  showstringspaces=false,
  breaklines=true,
  postbreak=\mbox{\textcolor{red}{$\hookrightarrow$}\space},
  numbers=left,
  numberstyle=\color{gray},
  stepnumber=1,
  firstnumber=1,
  numbersep=5pt,
}

\newcommand{\V}{\mathcal{V}}
\newcommand{\E}{\mathcal{E}}

\renewcommand{\S}{\mathcal{S}}

\newcommand{\R}{\mathcal{R}}

\renewcommand{\O}{\mathcal{O}}

\renewcommand{\O}{\mathcal{O}}

\newcommand{\T}{\mathcal{T}}

\newcommand{\bm}[1]{\text{\boldmath $#1$}}

\algdef{SE}[DOWHILE]{Do}{doWhile}{\algorithmicdo}[1]{\algorithmicwhile\ #1}%

\usepackage[usenames,dvipsnames,svgnames,table]{xcolor}
\usepackage{comment}

\newcommand{\Fin}{\color{black}}

\newif\iftoc
\toctrue
\usepackage{CJKutf8}

\def\BibTeX{{\rm B\kern-.05em{\sc i\kern-.025em b}\kern-.08em
T\kern-.1667em\lower.7ex\hbox{E}\kern-.125emX}}

\begin{document}
%
\title{Optimization of Model Splitting, Placement, and Chaining for Multi-hop Split Learning and Inference}
%
%
%

\author{Takanori~Hara,~\IEEEmembership{Member,~IEEE,}
  and~Masahiro~Sasabe,~\IEEEmembership{Member,~IEEE}
  \thanks{This work was supported in part by the Japan Society for the Promotion of Science (JSPS) KAKENHI (B) under Grant 25K03114 and 26K02903 and the Support Center for Advanced Telecommunications Technology Research (SCAT).}
\thanks{T.~Hara is with the Division of Information Science, Nara Institute of Science and Technology, Nara, Japan, e-mail: hara@ieee.org. M.~Sasabe is with the Faculty of Informatics, Kansai University, Takatsuki, Japan, email: m-sasabe@ieee.org.}}

\maketitle

\begin{abstract}
  Service Function Chaining (SFC) establishes efficient communication paths by ensuring that traffic traverses a predefined sequence of network functions in a specified order to meet particular service requirements.
  Inspired by this concept, we have proposed an SFC-based architecture for multi-hop split learning (MSL) and split inference (MSI), facilitating distributed AI applications to effectively route smashed data across multi-hop networks. 
  However, the multi-hop environment presents new challenges, including (1) determining optimal cut points, (2) deploying split sub-models on appropriate computing nodes, and (3) routing smashed data through the underlying communication networks while adhering to service requirements.
  To address these challenges, we formulate an Integer Linear Programming (ILP) model to jointly optimize model splitting, placement, and chaining (data routing) in the SFC-based MSL/MSI architecture, aiming to minimize end-to-end inference or training latency.
  Additionally, we propose a Block Coordinate Descent (BCD)-based heuristic algorithm to efficiently solve the problem.
  Comprehensive evaluations demonstrate the effectiveness and characteristics of the proposed formulation and algorithm.
\end{abstract}

\section{Introduction}
\label{sec:Introduction}

Recently, there has been growing demand for distributed Artificial Intelligence (AI) under stringent constraints on privacy, latency, power consumption, and computational resources through distributed execution across cloud-edge-end environments.
Split learning and inference (SL/SI) emerge as key enablers of distributed AI architecture by partitioning a global model into two sub-models, where one sub-model is deployed on a server and the other is deployed on multiple resource-constrained client devices~\cite{vepakommaSplitLearningHealth2018,zhuESFLEfficientSplit2024,linEfficientParallelSplit2024}.
Multi-hop split learning and inference (MSL/MSI) extend SL/SI to multi-hop network environments by partitioning a global model into multiple disjoint sub-models and deploying them across multiple computing nodes~\cite{tiranaEstimatingTrainingTime2025,linHierarchicalSplitFederated2025,haraServiceFunctionChaining2025,xuInferenceRoutingMultiHop2026,weiPipeliningSplitLearning2025}.
Unlike SL/SI, which typically assumes direct client-server communication, MSL/MSI must address the routing of processed data, commonly referred to as \textit{smashed data}, across multiple sub-models through multi-hop paths in the underlying communication networks.
This routing directly affects end-to-end performance metrics such as training and inference latency.

Service Function Chaining (SFC) establishes efficient communication paths by ensuring that traffic traverses a predefined sequence of network functions in a designated order to meet specific service requirements~\cite{rfc7665}.
In \cite{haraServiceFunctionChaining2025}, we proposed an SFC-based distributed AI architecture in which split sub-models are treated as network functions and their composition forms a service chain representing the global model.
However, this architecture introduces new challenges: (1) selecting optimal split points, (2) deploying split sub-models on appropriate computing nodes, and (3) routing smashed data through the underlying communication networks while satisfying service requirements.

To address these challenges, we formulate a joint optimization problem for model splitting, placement, and chaining (data routing) in the MSL/MSI architecture as an Integer Linear Programming (ILP) model.
The objective is to minimize training or inference latency while satisfying the required model execution order and resource constraints.
To solve the problem with lower computational cost, we develop a Block Coordinate Descent (BCD)-based algorithm that alternately optimizes (1) model splitting and (2) model placement and chaining.
Through extensive evaluation, we quantify the computing-communication tradeoff across mini-batch sizes and service chain lengths and demonstrate that the proposed heuristic achieves comparable latency to the ILP solution with significantly improved scalability.

The remainder of this paper is organized as follows.
Section~\ref{sec:Related Work} reviews related work.
Section~\ref{sec:System Model} describes the system model of the SFC-based MSL/MSI architecture.
Section~\ref{sec:Integer Programming} formulates the model splitting, placement, and chaining problem as an ILP model.
Section~\ref{sec:Heuristic Algorithm} proposes a heuristic algorithm to efficiently solve the model splitting, placement, and chaining problem.
Section~\ref{sec:Evaluation} presents evaluation results.
Section~\ref{sec:Conclusion} concludes the paper.

\section{Related Work}
\label{sec:Related Work}

Numerous studies have addressed resource allocation in model splitting and placement~\cite{yanOptimalModelPlacement2022, wuSplitLearningWireless2023, kimBargainingGamePersonalized2023, zhuESFLEfficientSplit2024,linEfficientParallelSplit2024,liAdaptiveSplitLearning2024,marinovaOptimalCutLayer2025,tiranaEstimatingTrainingTime2025,linHierarchicalSplitFederated2025}.
These studies formulate various optimization problems to minimize training/inference latency and energy consumption by optimizing model splitting and placement.
However, most existing approaches do not adequately address routing smashed data through the underlying communication network, which is critical in multi-hop environments.
Assuming that the model has already been split, Jung et al.\ formulated an ILP model for model placement and data routing to minimize end-to-end latency, including both computation and communication delays, using a layered graph~\cite{jungOptimizationFrameworkSplitting2023}.
Xu et al.\ proposed an optimization framework that jointly optimizes model splitting, placement, and data routing to minimize inference latency~\cite{xuInferenceRoutingMultiHop2026}.
However, their data routing approach relies on selecting from a predetermined set of route candidates.
In \cite{sartzetakisEdgeCloudInfiniteTime2024}, Sartzetakis et al.\ formulated a network and computation resource allocation problem for distributed machine learning as an ILP model to optimize monetary cost and accuracy while satisfying resource, accuracy, and latency requirements.
In \cite{tajiriOptimizationDataModel2025}, Tajiri et al.\ proposed a framework for optimizing data and model transfers in federated learning, formulated as a linear programming model to minimize the maximum aggregated data sent to the server.
Similarly, in \cite{fanDynamicTopologyResource2025}, Fan et al.\ proposed a dynamic virtual network embedding algorithm tailored for distributed training in mobile edge computing.
While these studies focus on optimizing either data routing or placement, they do not address the joint optimization of model splitting, placement, and chaining in the context of MSL/MSI.

To address these limitations, we formulate the joint optimization of model splitting, placement, and chaining as an ILP problem with the objective of minimizing inference or training latency.
The proposed ILP is an extension of our previous work~\cite{sasabeCapacitatedShortestPath2021}, which modeled an SFC and function placement problem as a capacitated shortest path tour problem (CSPTP) using an augmented network model.
CSPTP aims at finding the shortest path tour from a source node to a destination node while visiting at least one node from each of the given disjoint node subsets $\T_1, \ldots, \T_K$ in the specified order and satisfying the resource constraints.
In \cite{haraServiceFunctionChaining2025}, we emphasized the similarity between SFC and MSL/MSI and proposed an SFC-based MSL/MSI architecture.
In this paper, we extend the CSPTP formulation to jointly optimize model splitting, placement, and chaining in the SFC-based MSL/MSI architecture.

\begin{figure}[!t]
  \centering
  \includegraphics[width=0.9\columnwidth]{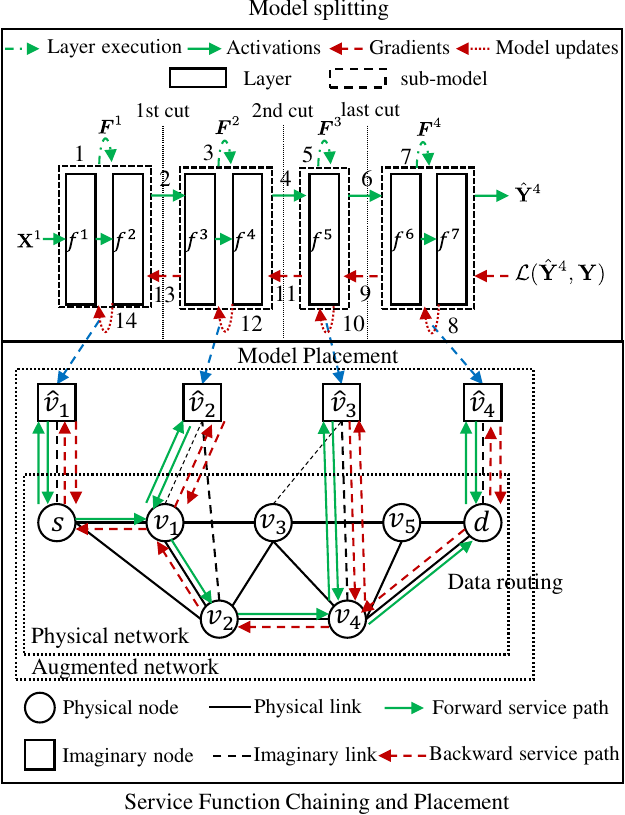}
  \caption{System model of SFC-based MSL/MSI.}
  \label{fig:system_model}
\end{figure}
\section{System Model}
\label{sec:System Model}
This section introduces the system model of the SFC-based MSL/MSI architecture, as illustrated in Fig.~\ref{fig:system_model}.

\subsection{Service Chain Request}

A user's request for executing a global model $\bm{F}$ is represented as a service chain request (SCR) $\R = (id, s, d, b, mode)$, where $id$ is the identifier of global model $\bm{F}$, $s$ is the source node, $d$ is the destination node, $b$ is the batch size, and $mode \in \{\mathrm{IF}, \mathrm{TR}\}$ represents the execution mode, i.e., inference $(\mathrm{IF})$ or training $(\mathrm{TR})$.

\subsection{Model Splitting}
\label{sec:Model Splitting}

A global model $\bm{F}$ consists of $L$ layers ($L=|\bm{F}|$).
The system partitions $\bm{F}$ into a sequence of $K \geq 2$ sub-models, $\bm{F}^k$ ($k=1,\ldots,K$).
Each sub-model $\bm{F}^k=(f^{s^k}, \ldots, f^{s^k+L^k-1})$ contains $L^k$ layers, where $L = \sum_{k=1}^{K}L^k$ and $L^k=|\bm{F}^k|$.
Here, $f^{l}$ represents the operation of the $l$th layer, and $s^k$ indicates the starting layer index of the $k$th sub-model $\bm{F}^k$, with $s^1=1$ and $s^k=s^{k-1}+L^{k-1}$ for $k \geq 2$.
In Fig.~\ref{fig:system_model}, $K = 4$ and $(L^1, L^2, L^3, L^4) = (2, 2, 1, 2)$.

During the inference phase, only forward computation (propagation) is performed, whereas the training phase involves both forward and backward computation.
During the forward computation, given an input matrix $\mathbf{X}^1$, the first sub-model $\bm{F}^1$ generates the activation matrix $\hat{\mathbf{Y}}^{L^1}$ (step 1 in Fig.~\ref{fig:system_model}) and transmits it to the next sub-model $\bm{F}^2$ (step 2 in Fig.~\ref{fig:system_model}).
Upon receiving the activation $\mathbf{X}^{s^k}$ from the preceding sub-model $\bm{F}^{k-1}$, the sub-model $\bm{F}^k$ computes the activations $\hat{\mathbf{Y}}^{s^k+L^k-1}$ (steps 3 and 5 in Fig.~\ref{fig:system_model}) and forwards them to the subsequent sub-model $\bm{F}^{k+1}$ (steps 4 and 6 in Fig.~\ref{fig:system_model}).
Finally, the last sub-model $\bm{F}^K$ computes the inference result $\hat{\mathbf{Y}}^{L}$ (step 7 in Fig.~\ref{fig:system_model}).

In the backward propagation phase, the last sub-model $\bm{F}^K$ calculates the loss function $\mathcal{L}(\hat{\mathbf{Y}}^{L}, \mathbf{Y})$, where $\mathbf{Y}$ represents the ground truth matrix.
It then computes the gradients, updates the weight parameters (step 8 in Fig.~\ref{fig:system_model}), and transmits the gradients to the preceding sub-model $\bm{F}^{K-1}$ (step 9 in Fig.~\ref{fig:system_model}).
Upon receiving the gradients from the succeeding sub-model $\bm{F}^{k+1}$, the sub-model $\bm{F}^k$ computes the gradients, updates the weight parameters (steps 10 and 12 in Fig.~\ref{fig:system_model}), and transmits the gradients to the preceding sub-model $\bm{F}^{k-1}$ (steps 11 and 13 in Fig.~\ref{fig:system_model}).
Finally, the first sub-model $\bm{F}^1$ computes the gradients and updates the weight parameters (step 14 in Fig.~\ref{fig:system_model}).

In the model splitting, both the number $K$ of sub-models and the split points $L^k$ $(k = 1,\ldots,K)$ must be carefully determined to minimize both computation and communication overheads.

\subsection{Physical Network and Augmented Network}
\label{sec:Physical Network and Augmented Network}

A physical network is represented as a directed graph $\mathcal{G}=(\V, \E)$, where $\V$ denotes the set of physical nodes and $\E$ represents the set of physical links.
Each physical node $i \in \V$ is equipped with either GPU or CPU resources for data processing.
Additionally, each physical node $i \in \V$ has memory and storage capacities denoted as $C_i^\mathrm{mem}$ and $C_i^\mathrm{disk}$, respectively.
Each physical link $(i,j) \in \E$ is characterized by its uplink (forward-direction) and downlink (backward-direction) bandwidths, $R_{i,j}^\mathrm{FW}$ and $R_{i,j}^\mathrm{BW}$, respectively, as well as its forward-direction and backward-direction propagation delays $d^\mathrm{FW}_{i,j}$ and $d^\mathrm{BW}_{i,j}$.

An augmented network $\mathcal{G}^+=(\V^+, \E^+)$ is an extension of the physical network, where $\V^+ = \V \cup \hat{\V}$ and $\E^+ = \E \cup \hat{\E}$~\cite{sasabeCapacitatedShortestPath2021}.
Here, $\hat{\V}$ and $\hat{\E}$ represent sets of imaginary nodes and links, respectively.
Each imaginary node $\hat{v}_k \in \hat{\V}$ corresponds to the sub-model $\bm{F}^k$ ($k=1,\ldots,K$).
The physical node executing $\bm{F}^k$ is selected from the set $\V^k \subseteq \V$, which represents the candidate nodes capable of hosting the sub-model $\bm{F}^k$.
Links $(i, \hat{v}_k)$ and $(\hat{v}_k, i)$ $(\hat{v}_k \in \hat{\V}, i \in \V^k)$ are called imaginary links, which represent the possibility of executing $\bm{F}^k$ on physical node $i$.
Hereafter, we use the notation $\mathcal{G}$ (resp.\ $\mathcal{G}^+$) to refer to the physical (resp.\ augmented) network including all associated node and link attributes when no confusion arises.

In Fig.~\ref{fig:system_model}, the imaginary nodes $\hat{v}_1$, $\hat{v}_2$, $\hat{v}_3$, and $\hat{v}_4$ represent the sub-models $\bm{F}^1$, $\bm{F}^2$, $\bm{F}^3$, and $\bm{F}^4$, respectively.
The imaginary links indicate that $\bm{F}^1$ and $\bm{F}^4$ are deployed on the origin and destination nodes $s$ and $d$, respectively, and that $\V^2 = \{v_1, v_2\}$ and $\V^3 = \{v_3, v_4\}$.

\subsection{Service Path}

Finding an optimal service path in the augmented network $\mathcal{G}^+$ corresponds to jointly optimizing model placement and chaining.
A service path $\S$ is composed of a sequence $(\S_1, \ldots, \S_{K+1})$ of $K+1$ subpaths.
Each $k$th subpath $\S_k$ connects its source and destination nodes $(a_k, b_k)$, where $a_k$ and $b_k$ are defined as follows: the first subpath starts at $s$ and ends at $\hat{v}_1$, intermediate subpaths connect consecutive imaginary nodes $\hat{v}_{k-1}$ and $\hat{v}_k$, and the last subpath connects $\hat{v}_K$ to $d$.
While the entire service path may include loops, each subpath $\S_k$ is loop-free.
Selecting an incoming imaginary link $(i, \hat{v}_k)$ in the $k$th subpath $\S_k$ indicates that the sub-model $\bm{F}^k$ is deployed and executed on the physical node $i$.
The cost of an imaginary link is determined by the forward and backward computation delays, while the cost of a physical link is determined by the activation and gradient transmission delays, as well as the link propagation delay.

In the forward computation illustrated in Fig.~\ref{fig:system_model}, the forward service path consists of five subpaths: $\S_1=(s, \hat{v}_1)$, $\S_2=(\hat{v}_1, s, v_1, \hat{v}_2)$, $\S_3=(\hat{v}_2, v_1, v_2, v_4, \hat{v}_3)$, $\S_4=(\hat{v}_3, v_4, d, \hat{v}_4)$, and $\S_5=(\hat{v}_4, d)$.
Therefore, the sub-models $\bm{F}_1$, $\bm{F}_2$, $\bm{F}_3$, and $\bm{F}_4$ are deployed on the physical nodes $s$, $v_1$, $v_4$, and $d$, respectively.
In the backward computation depicted in Fig.~\ref{fig:system_model}, the backward service path is the reverse of the forward one.

\section{Integer Linear Programming for Model Splitting, Placement, and Chaining}
\label{sec:Integer Programming}

We formulate the model splitting, placement, and chaining problem $\mathsf{P}_\mathrm{IF}$ (resp.\ $\mathsf{P}_\mathrm{TR}$) for the SFC-based MSI (resp.\ MSL) architecture as an extension of CSPTP-based ILP~\cite{sasabeCapacitatedShortestPath2021}:
\begin{align}
  \min
  \ 
  &
  T(\bm{x}, \bm{y}, b, \mathrm{FW}) + \mathbb{I}(mode = \mathrm{TR}) \cdot T(\bm{x}, \bm{y}, b, \mathrm{BW})
  \label{eq:objective_function}
  \\
  \text{s.t.}
  \ 
  &
  x_{i,j}^k \in \{0, 1\},
  \quad
  \forall (i, j) \in \E^+, \forall k \in \{1, \ldots, K+1\},
  \label{con:binary_variable:x}
  \\
  &
  \sum_{j \in \V_i^+} x_{i,j}^k - \sum_{j \in \V_i^+} x_{j,i}^k =
  \begin{cases}
    1 & \textrm{if $i = a_{k}$,}\\
    -1 & \textrm{if $i = b_{k}$,}\\
    0 & \textrm{otherwise,}\\
  \end{cases}
  \nonumber
  \\
  &
  \quad
  \forall i \in \V^+ \setminus \{a_k, b_k\}, \forall k \in \{1, \ldots, K+1\},
  \label{con:flow_conservation}
  \\
  &
  x_{i,\hat{v}_k}^k = x_{\hat{v}_k,i}^{k+1},
  \quad
  \forall (i, \hat{v}_k) \in \E^+, \forall k \in \{1, \ldots, K+1\},
  \label{con:ensure_connectivity}
  \\
  &
  x_{i,\hat{v}_m}^k = 0,
  \quad
  \forall (i, \hat{v}_m) \in \E^+, \forall k \in \{1, \ldots, K+1\},
  \label{con:prohibit_invalid_connectivity}
  \\
  &
  y_{\hat{v}_k,l} = \{0, 1\},
  \quad
  \forall \hat{v}_k \in \hat{\V}, \forall l \in \{1, \ldots, L\},
  \label{con:binary_variable:mu}
  \\
  &
  y_{\hat{v}_1, 1} = 1,
  \label{con:first_layer}
  \\
  &
  y_{\hat{v}_K, L} = 1,
  \label{con:last_layer}
  \\
  &
  \sum_{\hat{v}_k \in \hat{\V}} y_{\hat{v}_k,l} = 1,
  \quad
  \forall l \in \{1, \ldots, L\},
  \label{con:one_layer_is_assigned_to_node}
  \\
  &
  \sum_{l=1}^L y_{\hat{v}_k,l} \geq 1,
  \quad
  \forall \hat{v}_k \in \hat{\V},
  \label{con:imaginary_node_holds_one_or_more_layer}
  \\
  &
  y_{\hat{v}_1,0} = 0,
  \label{con:dummy}
  \\
  &
  \sum_{l=k}^L \max(0, y_{\hat{v}_k,l} - y_{\hat{v}_k,l-1}) = 1,
  \quad
  \forall k \in \{1, \ldots, K\},
  \label{con:ensure_sequent_layers}
  \\
  &
  y_{\hat{v}_k,l} - y_{\hat{v}_k,l-1} \leq y_{\hat{v}_{k-1},l-1},
  \nonumber
  \\
  &
  \quad
  \forall k \in \{2, \ldots, K\}, \forall l \in \{2, \ldots, L\},
  \label{con:ensure_sequent_layers3}
  \\
  &
  x_{i,\hat{v}_k}^k \sum_{l=1}^L y_{\hat{v}_k,l}r_l^\mathrm{disk} \leq C_i^\mathrm{disk},
  \quad
  \forall (i, \hat{v}_k) \in \E^+,
  \label{con:storage_constraint}
  \\
  &
  x_{i,\hat{v}_k}^k \sum_{l=1}^L y_{\hat{v}_k,l}r_l^\mathrm{mem} + b\max_{\substack{l \in \{1,\ldots, L\},\\ dir \in D(mode)}}
  x_{i,\hat{v}_k}^k y_{\hat{v}_k,l} \delta_l^{dir} \leq C_i^\mathrm{mem},
  \nonumber
  \\
  &
  \quad
  \forall (i, \hat{v}_k) \in \E^+.
  \Fin
  \label{con:memory_constraint}
\end{align}

Here, there are two kinds of decision variables.
Let $\bm{x} = [x_{i,j}^{k}]$ $(k \in \{1, \ldots, K+1\}, (i,j) \in \E^+)$ denote the binary decision variables for model placement and chaining:
\[
  x_{i,j}^{k} =
  \begin{cases}
    1,& \text{if physical/imaginary link $(i,j)$ is included in the}\\
      & \text{$k$th subpath of the service path},\\
    0,& \text{otherwise}.
  \end{cases}
\]
Let $\bm{y} = [y_{\hat{v}_k,l}]$ $(\hat{v}_k \in \hat{\V}, l \in \{1, \ldots, L\})$ denote the binary decision variables for model splitting:
\[
  y_{\hat{v}_k,l} =
  \begin{cases}
    1,& \text{if the $l$th layer is assigned to imaginary node $\hat{v}_k$},\\
    0,& \text{otherwise}.
  \end{cases}
\]

The difference between $\mathsf{P}_\mathrm{IF}$ and $\mathsf{P}_\mathrm{TR}$ is reflected in the objective function~\eqref{eq:objective_function}, which minimizes the total inference latency $T(\bm{x}, \bm{y}, \mathrm{IF}) = T(\bm{x}, \bm{y}, b, \mathrm{FW})$ in the inference case or the total training latency $T(\bm{x}, \bm{y}, \mathrm{TR}) = T(\bm{x}, \bm{y}, b, \mathrm{FW}) + T(\bm{x}, \bm{y}, b, \mathrm{BW})$ in the training case.
In Eq.~\eqref{eq:objective_function}, $\mathbb{I}(\cdot)$ is the indicator function, which equals 1 if the condition is satisfied and 0 otherwise.
Here, $T(\bm{x}, \bm{y}, b, dir)$ is defined as the sum of the total computation (processing) delay and the total communication delay for the forward ($dir = \mathrm{FW}$) or backward ($dir = \mathrm{BW}$) direction, which is expressed as follows:
\begin{align}
  T(\bm{x}, \bm{y}, b, dir)
  &=
  \sum_{(i, \hat{v}_k) \in \hat{\E}} x_{i,\hat{v}_k}^k T_{i,\hat{v}_k}^\mathrm{comp}(\bm{y}, b, dir)
  \nonumber
  \\
  &+
  \sum_{(\hat{v}_k, m) \in \hat{\E}} x_{\hat{v}_k,m}^{k+1}
  \sum_{(i,j) \in \E} x_{i,j}^{k+1}
  \left(
  T_{i,j,\hat{v}_k}^\mathrm{trans}(\bm{y}, b, dir)
  +
  d^{dir}_{i,j}
  \right)
  \label{eq:total_latency}
\end{align}
The first term consists of the forward/backward computation delay $T_{i,\hat{v}_k}^\mathrm{comp}(\bm{y}, b, dir)$ for executing a sub-model $\bm{F}^k$ $(k = 1, \ldots, K)$ at a physical node $i$, which is defined as
\begin{align}
  T_{i,\hat{v}_k}^\mathrm{comp}(\bm{y}, b, dir) = \kappa_i(b, \phi_{\hat{v}_k}(\bm{y}, \bm{\rho}^{dir})) + \tau_i(b).
  \label{eq:forward_propagation_latency}
\end{align}
Recall $b$ denotes the batch size.
$\bm{\rho}^{dir} = [\rho_1^{dir}, \ldots, \rho_L^{dir}]$ represents the vector of Floating-point Operations (FLOPs),
which are required for the forward ($dir = \mathrm{FW}$) or backward ($dir = \mathrm{BW}$) computation of each layer.
$\kappa_i()$ and $\tau_i()$ are defined as linear functions that derive the computing time for executing a sub-model $\bm{F}^k$ at physical node $i$ and that output a GPU I/O overhead of physical node $i$, respectively:
\begin{align}
  \kappa_i(b, \phi) = (\alpha_{\kappa} \cdot b + \beta_{\kappa}) \cdot \phi,
  \quad
  \tau_i(b) = \alpha_{\tau} \cdot b + \beta_{\tau}.
  \nonumber
\end{align}
Here, $\alpha_{\kappa}$, $\beta_{\kappa}$, $\alpha_{\tau}$, and $\beta_{\tau}$ are constant values, which would be fitted by ordinary least squares (OLS) using actual measurements of computing nodes.
In case of the CPU node, $\tau_i(b)$ is always 0, so $\alpha_{\tau} = \beta_{\tau} = 0$.
The forward/backward computation workload $\phi_{\hat{v}_k}(\bm{y}, \bm{\rho}^{dir})$ is the workload to execute the assigned layers:
\begin{align}
  \phi_{\hat{v}_k}(\bm{y}, \bm{\rho}^{dir}) = \sum_{l=1}^{L} y_{\hat{v}_k,l} \rho_l^{dir}.
  \nonumber
\end{align}

The total forward (resp.\ backward) communication delay consists of the total activation (resp.\ gradient) transmission delay and the total forward-direction (resp.\ backward-direction) link propagation delay.
The total activation/gradient transmission delay consists of the activation/gradient transmission delay $T_{i,j,\hat{v}_k}^\mathrm{trans}(\bm{y}, b, dir)$ between link $(i,j)$ in the $(k+1)$th subpath $(k = 1, \ldots, K)$, which is defined as
\begin{align}
  T_{i,j,\hat{v}_k}^\mathrm{trans}(\bm{y}, b, dir)
  &=
  \frac{b \psi_{\hat{v}_k}(\bm{y}, \bm{\delta}^{dir})}{R_{i,j}^{dir}},
  \label{eq:activation_transmission_latency}
\end{align}
where $\bm{\delta}^{dir} = [\delta_1^{dir}, \ldots, \delta_{L-1}^{dir}]$ represents the vector of the activation/gradient size of each layer in bytes, and $R_{i,j}^{dir}$ denotes the forward-direction/backward-direction bandwidth of physical link $(i,j)$.
Let $\psi_{\hat{v}_k}(\bm{y}, \bm{\delta}^{dir})$ denotes the activation/gradient size of $k$th sub-model:
\begin{align}
  \psi_{\hat{v}_k}(\bm{y}, \bm{\delta}^{dir}) = \sum_{l=1}^{L-1} y_{\hat{v}_k,l}(1 - y_{\hat{v}_k,l+1}) \delta_l^{dir}.
  \nonumber
\end{align}

The total forward-direction/backward-direction link propagation delay consists of the forward-direction/backward-direction link propagation delay $d^{dir}_{i,j}$ of link $(i,j)$ in the $(k+1)$th subpath $(k = 1, \ldots, K)$.

Constraints~\eqref{con:binary_variable:x}--\eqref{con:prohibit_invalid_connectivity} are associated with model placement and chaining, which are the same as those in the CSPTP-based ILP by regarding functions as models.
Constraint~\eqref{con:binary_variable:x} represents the binary decision variables.
Constraint~\eqref{con:flow_conservation} ensures the flow conservation rules of the service path, where $\V_i^+$ denotes a set of neighbor nodes of $i$.
Constraint~\eqref{con:ensure_connectivity} guarantees the connectivity between the $k$th and $k+1$th subpaths.
Constraint~\eqref{con:prohibit_invalid_connectivity} prohibits the service path from traversing the imaginary node $\hat{v}_m$ in the $k$th subpath ($m \neq k$).

\begin{figure}[!t]
  \centering
  \includegraphics[width=0.9\columnwidth]{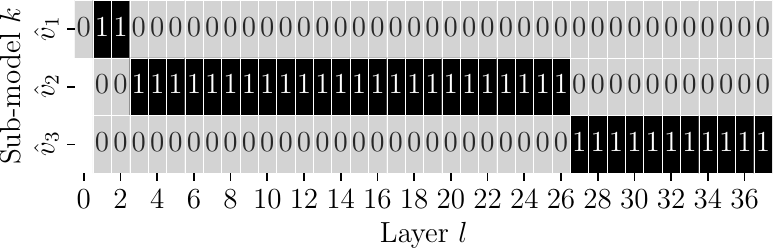}
  \caption{Relationship between $y_{\hat{v}_k,l}$ and valid model splitting ($K=3, L=37$).}
  \label{fig:valid_model_splitting}
\end{figure}
The remaining constraints~\eqref{con:binary_variable:mu}--\eqref{con:memory_constraint} are newly added or modified to the CSPTP-based ILP.
Constraints~\eqref{con:binary_variable:mu}--\eqref{con:ensure_sequent_layers3} ensure the valid model splitting.
Fig.~\ref{fig:valid_model_splitting} illustrates the relationship between $y_{\hat{v}_k,l}$ and valid model splitting for $K=3$ and $L=37$.
Constraint~\eqref{con:binary_variable:mu} represents the binary decision variables.
Constraints~\eqref{con:first_layer} and \eqref{con:last_layer} ensure that the first and last layers are assigned to the first imaginary node $\hat{v}_1$ and the last imaginary node $\hat{v}_K$, respectively.
Constraint~\eqref{con:one_layer_is_assigned_to_node} ensures that each layer is always assigned to one imaginary node.
Constraint~\eqref{con:imaginary_node_holds_one_or_more_layer} guarantees that each imaginary node executes at least one layer.
Under the definition of the dummy variable $y_{\hat{v}_1,0} = 0$ in constraint~\eqref{con:dummy}, constraint~\eqref{con:ensure_sequent_layers} guarantees that the layers assigned to each sub-model are contiguous.
As illustrated in Fig.~\ref{fig:valid_model_splitting}, for each $k$, the values of $y_{\hat{v}_k,l}$ must start at 0 for layers unassigned to sub-model $F^k$, transition to 1 for the consecutive layers assigned to sub-model $F^k$, and revert to 0 for the remaining layers.
In this scenario, $\max(0, y_{\hat{v}_k,l} - y_{\hat{v}_k,l-1})$ in the left-hand side of constraint~\eqref{con:ensure_sequent_layers} evaluates to 1 only at the layer $l$ where the assignment to sub-model $F^k$ begins, and 0 elsewhere.
The sum of these values is constrained to equal 1.
To ensure that the layers assigned to consecutive sub-models $F^{k-1}$ and $F^k$ are sequentially ordered, the relationship $y_{\hat{v}_{k-1},l-1} = 1$, $y_{\hat{v}_k,l-1} = 0$, and $y_{\hat{v}_k,l} = 1$ must hold at the first layer $l$ assigned to sub-model $F^k$.
For instance, in Fig.~\ref{fig:valid_model_splitting}, this corresponds to $y_{\hat{v}_1,2} = 1$, $y_{\hat{v}_2,2} = 0$, and $y_{\hat{v}_2,3} = 1$.
Here, the left-hand side of constraint~\eqref{con:ensure_sequent_layers3} evaluates to 0 until the assignment to sub-model $F^k$ begins, becomes 1 at the first assigned layer, remains 0 until the last assigned layer, evaluates to $-1$ at the next layer, and then reverts to 0.
Therefore, the combination of constraint~\eqref{con:ensure_sequent_layers} and constraint~\eqref{con:ensure_sequent_layers3} guarantees that the layers assigned to consecutive sub-models are sequentially ordered.

Constraint~\eqref{con:storage_constraint} ensures that the total size of the layers assigned to the sub-model $\bm{F}^k$ does not exceed the storage capacity $C_i^\mathrm{disk}$ of the physical node $i$ executing $\bm{F}^k$, where $r_l^\mathrm{disk}$ represents the disk size required to store the $l$th layer.
Constraint~\eqref{con:memory_constraint} enforces the memory capacity constraint, where $r_l^\mathrm{mem}$ denotes the memory size required for the $l$th layer, and $C_i^\mathrm{mem}$ represents the memory capacity of physical node $i$.
This constraint is similar to the storage constraint~\eqref{con:storage_constraint}.
However, during batch processing, activations or gradients must be temporarily stored in memory for each layer in the sub-model $\bm{F}^k$.
The maximum size of these temporary data is represented by the second term on the left-hand side.
Here, $D(mode)$ is defined as $\{\mathrm{FW}, \mathrm{BW}\}$ for $mode = \mathrm{TR}$, and as $\{\mathrm{FW}\}$ otherwise.
In practical scenarios, the memory constraint \eqref{con:memory_constraint} is often more stringent than the storage constraint \eqref{con:storage_constraint}, as storage systems typically employ compressed formats for model storage and incur lower costs per byte.
In practical scenarios, the memory constraint \eqref{con:memory_constraint} is typically more restrictive than the storage constraint \eqref{con:storage_constraint}, as storage systems often utilize compressed formarts for model strage and incur lower costs per byte.

Although the objective function~\eqref{eq:objective_function}, constraints~\eqref{con:ensure_sequent_layers}, \eqref{con:memory_constraint}, and \eqref{con:storage_constraint} are non-linear, due to the product of two or more binary decision variables and the max operator, these can be linearized using existing techniques~\cite{vanderbeiLinearProgramming2015}.

\section{Block Coordinate Descent-Based Heuristic Algorithm}
\label{sec:Heuristic Algorithm}

\subsection{Overview}

\begin{algorithm}[t]
  \caption{BCD-based heuristic algorithm.}
  \label{alg:bcd}
  \footnotesize
  \begin{algorithmic}[1]
  \Require $\R = (id, s, d, b, mode), \mathcal{G}, L, \{\V^k\}_{k=1,\ldots,K}, \bm{\rho}, \bm{\delta}, \varepsilon$.
    \Ensure $\bm{x}^*$, $\bm{y}^*$.
    \State $t \leftarrow 0$
    \label{alg:bcd:begin_initialization}
    \State Initialize $\bm{y}_0$
    \State $\mathcal{G}^+ \leftarrow \Call{AugmentedNetwork}{\mathcal{G}, \R, \bm{y}_0, \{\V^k\}_{k=1,\ldots,K}, \bm{\rho}, \bm{\delta}}$
    \State $\bm{x}_0 \leftarrow \Call{DFTS}{\mathcal{G}^+, s, d}$
    \label{alg:bcd:end_initialization}
    \Repeat
    \label{alg:bcd:begin_iteration}
    \State $t \leftarrow t + 1$
    \label{alg:bcd:update_t}
    \State $\bm{y}_{t} \leftarrow \Call{K-Sequence-Segmentation}{\bm{x}_{t-1}, \R, \mathcal{G}^+, L, \bm{\rho}, \bm{\delta}}$
    \label{alg:bcd:k-sequence-segmentation}
    \State $\mathcal{G}^+ \leftarrow \Call{AugmentedNetwork}{\mathcal{G}, \bm{y}_{t}, \R, \{\V^k\}_{k=1,\ldots,K}, \bm{\rho}, \bm{\delta}}$
    \State $\bm{x}_{t} \leftarrow \Call{DFTS}{\mathcal{G}^+, s, d}$
    \label{alg:bcd:dfts}
    \Until{$|T(\bm{x}_{t}, \bm{y}_{t}, mode) - T(\bm{x}_{t-1}, \bm{y}_{t-1}, mode)| \leq \varepsilon$}
    \label{alg:bcd:end_iteration}
  \end{algorithmic}
\end{algorithm}

To solve the proposed ILP more efficiently, we propose a heuristic algorithm based on the block coordinate descent (BCD) method~\cite{wrightCoordinateDescentAlgorithms2015}.
The algorithm decomposes the primary problem $\mathsf{P}_\mathrm{IF}$ or $\mathsf{P}_\mathrm{TR}$ into two subproblems: (1) model splitting and (2) model placement and chaining.
It then iteratively optimizes these subproblems until convergence.
Algorithm~\ref{alg:bcd} outlines the BCD-based heuristic algorithm.
For solving $\mathsf{P}_\mathrm{IF}$, the vectors $\bm{\rho}$ and $\bm{\delta}$ are set as $\bm{\rho}^\mathrm{FW}$ and $\bm{\delta}^\mathrm{FW}$, respectively.
For solving $\mathsf{P}_\mathrm{TR}$, $\bm{\rho}$ and $\bm{\delta}$ are set as $(\bm{\rho}^\mathrm{FW}, \bm{\rho}^\mathrm{BW})$ and $(\bm{\delta}^\mathrm{FW}, \bm{\delta}^\mathrm{BW})$, respectively.

In the initialization phase (lines~\ref{alg:bcd:begin_initialization}--\ref{alg:bcd:end_initialization}), the iteration counter $t$ is initialized to zero.
The initial model splitting $\bm{y}_0$ is determined by evenly dividing the global model into $K$ layers.
The augmented network $G^+$ is then constructed based on the physical network $G$, the service chain request $\R$, $\bm{y}_0$, the placement candidates $\{\V^k\}_{k=1,\ldots,K}$, and the vectors $\bm{\rho}$ and $\bm{\delta}$ by invoking the $\Call{AugmentedNetwork}$ function.
Using the augmented network $G^+$, the source node $s$, and the destination node $d$, the initial model placement and chaining $\bm{x}_0$ is derived using the depth-first tour search (DFTS) algorithm~\cite{bhatServiceConcatenationRoutingApplications2017} (i.e., the $\Call{DFTS}$ function), which identifies the shortest path tour from $s$ to $d$.

During each iteration (lines~\ref{alg:bcd:begin_iteration}--\ref{alg:bcd:end_iteration}), the algorithm alternates between optimizing $\bm{y}_{t}$ and $\bm{x}_{t}$.
First, $t$ is incremented by one (line~\ref{alg:bcd:update_t}).
The model splitting $\bm{y}_{t}$ is optimized using the $K$-sequence segmentation algorithm via dynamic programming (DP)~\cite{bellmanApproximationCurvesLine1961} (i.e., the $\Call{K-Sequence-Segmentation}$ function) for the given model placement and chaining $\bm{x}_{t-1}$, $\R$, the vectors $\bm{\rho}$ and $\bm{\delta}$, and the number $L$ of layers (line~\ref{alg:bcd:k-sequence-segmentation}).
Next, the model placement and chaining $\bm{x}_{t}$ is optimized using the DFTS algorithm for the updated model splitting $\bm{y}_{t}$ (line~\ref{alg:bcd:dfts}).
If the difference between the objective function values at iterations $t$ and $t-1$ is less than or equal to the convergence tolerance $\varepsilon \geq 0$ (line~\ref{alg:bcd:end_iteration}), the algorithm terminates and outputs the final solutions $\bm{x}^*$ and $\bm{y}^*$.

\subsection{$K$-Sequence Segmentation via Dynamic Programming}

\begin{algorithm}[t]
  \caption{$K$-Sequence Segmentation via DP.}
  \label{alg:kss}
  \footnotesize
  \begin{algorithmic}[1]
    \Require $\bm{x}_{t-1}, \R = (id, s, d, b, mode), \mathcal{G}^+, L, \bm{\rho}, \bm{\delta}$.
    \Ensure $\bm{y}_{t}$.
    \For {$l \in \{1, \ldots, L - K\}$}
    \label{alg:kss:begin_initialization}
    \State $dp_{1,l} \leftarrow T(\bm{x}_{t-1}^1, \bm{1}_{1,l, L}^1, b, mode)$
    \EndFor
    \label{alg:kss:end_initialization}
    \For {$k \in \{2, \ldots, K\}$}
    \label{alg:kss:begin_transition}
    \If {$k < K$}
    \State $\mathcal{L}' \leftarrow \{k, \ldots, L-K-k+2\}$
    \Else
    \State $\mathcal{L}' \leftarrow \{L\}$
    \Comment The last segment must include the last layer
    \EndIf
    \For {$l \in \mathcal{L}'$}
    \For {$l' \in \{k-1, \ldots l-1\}$}
    \State $dp_{k,l} \leftarrow \min(dp_{k,l}, dp_{k-1, l'} + T(\bm{x}_{t-1}^k, \bm{1}_{l',l, L}^k, b, mode))$
    \EndFor
    \EndFor
    \EndFor
    \label{alg:kss:end_transition}
  \end{algorithmic}
\end{algorithm}

$K$-sequence segmentation~\cite{bellmanApproximationCurvesLine1961} aims at dividing a sequence of length $L$ into $K$ segments to minimize the total cost of the segments.
In our problem, the sequence and segments correspond to the layers of the global model and the sub-models, respectively.
Algorithm~\ref{alg:kss} shows the $K$-sequence segmentation via DP to optimize the model splitting $\bm{y}_{t}$ for a given model placement and chaining $\bm{x}^{(t-1)}$.
The recurrence relation $dp_{k,l}$ for solving the $K$-sequence segmentation problem via DP is defined as follows:
\begin{align}
  dp_{k,l} = \min_{1 \leq l' < l} (dp_{k-1,l'} + T(\bm{x}^k, \bm{1}_{l',l, L}^k, b, mode)).
  \nonumber
\end{align}
Here, $\bm{x}^k = [x_{i,j}^{k}]$ $(i,j \in \E^+)$ determines $k$th subpath $\S_k$ $(k = 1, \ldots, K)$.
Let $\bm{1}_{l',l, L}^k$ denote a binary vector of length $L$, which has 1s from the $l'$th to the $(l-1)$th elements and 0s elsewhere ($1 \leq l' < l \leq L$).
For example, $\bm{1}_{2,4,5}^k = [0, 1, 1, 0, 0]$ indicates that layers 2, 3 are assigned to the $k$th sub-model.
$\bm{1}_{l',l, L}^k$ can be viewed as a specific instance of $\bm{y}_{\hat{v}_k} = [y_{\hat{v}_k,1}, \ldots, y_{\hat{v}_k,L}]$, i.e., assignment of layers to the $k$th sub-model.
$T(\bm{x}_{t-1}^k, \bm{1}_{l',l, L}^k, b, mode)$ indicates the total latency of executing the $k$th sub-model with the layer assignment $\bm{1}_{l',l, L}^k$ and the model placement and chaining $\bm{x}_{t-1}^k$.
Note that this value is set to infinity if the storage or memory capacity constraints of physical node $i$ are violated.

In lines~\ref{alg:kss:begin_initialization}--\ref{alg:kss:end_initialization}, $dp_{1,l}$ is initialized for the first segment.
In lines~\ref{alg:kss:begin_transition}--\ref{alg:kss:end_transition}, the algorithm iteratively minimizes the cost $dp_{k,l}$ by evaluating all feasible values of $l$ for each segment $k$, processing sequentially from $k = 2$ to $K$.
Finally, the algorithm derives the optimal solution $\bm{y}$ by backtracking from $dp_{K,L}$.

\subsection{Shortest Path Tour Finding}
\label{sec:Shortest Path Tour Finding}

To optimize the model placement and chaining $\bm{x}$, we use the DFTS algorithm~\cite{bhatServiceConcatenationRoutingApplications2017}, which identifies the shortest path tour from the source node $s$ to the destination node $d$ while ensuring that all imaginary nodes $\hat{v}_k \in \hat{\V}$ are visited in the required order.
Inspired by our previous work on the VNF placement and chaining problem~\cite{haraSpeedyEfficientService2022}, we construct a modified augmented network $\mathcal{G}^+$ from the physical network $\mathcal{G}$, utilizing a given $\bm{y}$ to ensure connectivity between the $k$th and $k+1$th subpaths in the DFTS algorithm.\footnote{For simplicity of notation, we use the same symbol $\mathcal{G}^+$ as the original augmented network, provided there is no risk of confusion.}
This modified augmented network $\mathcal{G}^+$ differs slightly from the original augmented network described in Section~\ref{sec:Physical Network and Augmented Network}.

In the modified augmented network $\mathcal{G}^+$, the set of imaginary nodes is updated as $\hat{\V} = \{\hat{v}_{i,k}\}_{i \in \V^k, k \in \{1, \ldots, K\}}$, where each imaginary node $\hat{v}_{i,k}$ represents a pair of a placement candidate physical node $i \in \V^k$ and the split sub-model $\bm{F}^k$.
The corresponding imaginary links are also updated as $\{(\hat{v}_{i,k}, i)\}_{i \in \V^k, k \in \{1, \ldots, K\}}$ and $\{(i, \hat{v}_{i,k})\}_{i \in \V^k, k \in \{1, \ldots, K\}}$. 

The link cost $c_{i,j}^k$ of physical link $(i,j)$ in the $k$th subpath for forward/backward communication is defined as:
\begin{align*}
  c_{i,j}^k 
  &=
  T_{i,j,\hat{v}_{m,k-1}}^\mathrm{trans}(\bm{y}, b, \mathrm{FW}) + d_{i,j}^\mathrm{FW} \\
  &\quad + \mathbb{I}(mode = \mathrm{TR}) \cdot \bigl(T_{i,j,\hat{v}_{m,k-1}}^\mathrm{trans}(\bm{y}, b, \mathrm{BW}) + d_{i,j}^\mathrm{BW}\bigr),
\end{align*}
where $\hat{v}_{m,k} \in \hat{\V}$ is the imaginary node connected to the physical node $m$ that holds the $k$th sub-model $\bm{F}^k$.

The link cost $c_{i,\hat{v}_{i,k}}^k$ of an imaginary link $(i, \hat{v}_{i,k})$ for forward/backward computation is defined as:
\begin{align*}
  c_{i,\hat{v}_{i,k}}^k 
  &=
  T_{i,\hat{v}_{i,k}}^\mathrm{comp}(\bm{y}, b, \mathrm{FW}) \\
  &\quad + \mathbb{I}(mode = \mathrm{TR}) \cdot T_{i,\hat{v}_{i,k}}^\mathrm{comp}(\bm{y}, b, \mathrm{BW}).
\end{align*}

If the storage or memory capacity constraints of physical node $i$ are violated, the corresponding imaginary link is not created.
By applying the DFTS algorithm to the modified augmented network $\mathcal{G}^+$, along with the source node $s$ and destination node $d$, we can derive the shortest path tour and the model placement results (i.e., $\bm{x}$).

\subsection{Computational Complexity and Convergence Property}
\label{sec:Computational Complexity}

We first evaluate the computational complexity of the BCD-based heuristic algorithm.
The computational complexity of the DFTS algorithm is $\O((K+1)V^+)+\O((K+1)E^+\log V^+)$~\cite{haraSpeedyEfficientService2022}, where the first term represents the complexity of calculating the minimum cost for each subpath, and the second term corresponds to the complexity of the Dijkstra algorithm for determining the entire shortest path.
The computational complexity of the $K$-sequence segmentation via DP is $\O(K L^2)$~\cite{bellmanApproximationCurvesLine1961}.
Assuming that the DFTS algorithm and the $K$-sequence segmentation algorithm are invoked at most $T_{\max}+1$ and $T_{\max}$ times, respectively, the overall computational complexity of the BCD algorithm is $\O(T_{\max}K L^2) + \O((T_{\max}+1)(K+1)V^+) + \O((T_{\max} + 1)(K+1)E^+\log V^+)$.

In theory, BCD algorithms are not guaranteed to converge to a global optimal solution for ILP problems due to a non-convex nature~\cite{grippoConvergenceBlockNonlinear2000}.
However, it is well-known that the BCD algorithm often converges to a good solution in practice~\cite{linEfficientParallelSplit2024}.

\begin{figure}[!t]
  \centering
  \includegraphics[width=0.65\columnwidth]{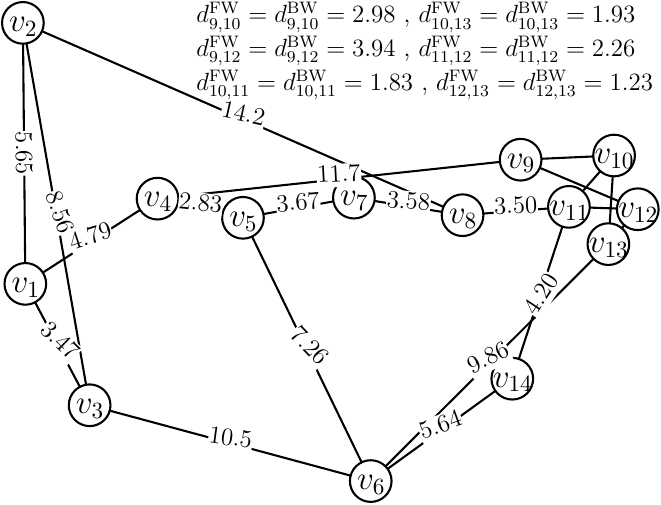}
  \caption{NSFNET topology~\cite{millsNSFNETBackboneNetwork1987}.}
  \label{fig:nsfnet}
\end{figure}

\begin{table*}[!t]
  \caption{Model specification of ResNet101~\cite{heDeepResidualLearning2016} ($3\times 224 \times 224$-dimension input with $b=1$).}
  \label{table:model_specification}
  \centering
  \begin{tblr}{
    colspec={llrrrrr},
    hline{1,3,Z} = {solid},
    cell{1}{3} = {c=2}{c},
    cell{1}{5} = {c=2}{c},
  }
    Building block & Building block name & Computational cost (FLOPs) & & Smashed data size (bytes) & & Layer size (bytes) \\
    (Layer) id $l$ & & Forward $\rho_l^\mathrm{FW}$ & Backward $\rho_l^\mathrm{BW}$ & Activation $\delta_l^\mathrm{FW}$ & Gradient $\delta_l^\mathrm{BW}$ & $r_l^\mathrm{mem}, r_l^\mathrm{disk}$ \\
    1      & conv1   & 236.02\,M & 472.04\,M & 3.21\,M & 3.21\,M & 37\,K \\
    2      & conv2\_x (batchnorm, relu, maxpool) & 6.43\,M & 12.9\,M & 0.80\,M & 0.80\,M & 512 \\
    3      & conv2\_x   & 4.74\,G & 9.48\,G & 3.21\,M & 3.21\,M & 3.02\,M \\
    4,5    & conv2\_x   & 7.40\,G & 14.80\,G & 3.21\,M & 3.21\,M & 4.72\,M \\
    6      & conv3\_x   & 5.76\,G & 11.52\,G & 1.61\,M & 1.61\,M & 14.68\,M \\
    7--9   & conv3\_x   & 7.40\,G & 14.80\,G & 1.61\,M & 1.61\,M & 18.88\,M \\
    10     & conv4\_x   & 5.76\,G & 11.52\,G & 0.80\,M & 0.80\,M & 58.76\,M \\
    11--32 & conv4\_x   & 7.40\,G & 14.80\,G & 0.80\,M & 0.80\,M & 75.52\,M \\
    33     & conv5\_x   & 5.76\,G & 11.52\,G & 0.40\,M & 0.40\,M & 234.92\,M \\
    34-35  & conv5\_x   & 7.40\,G & 14.80\,G & 0.40\,M & 0.40\,M & 302.04\,M \\
    36     & avgpool    & 200.70\,K & 401.40\,K & 8192 & 8192 & 0 \\
    37     & fc         & 4.10\,M & 8.20\,M & 4000 & 4000 & 8.20\,M \\
  \end{tblr}
\end{table*}

\begin{table}[!t]
  \caption{Computing nodes' specifications.}
  \label{table:computing_node_specification}
  \footnotesize
  \centering
  \newcolumntype{L}{>{\raggedright\arraybackslash}X}
  \begin{tabularx}{\columnwidth}{lll}
    \toprule
    Node type         & CPU & GPU \\
         & Intel(R) Xeon(R) Gold 6226R &  NVIDIA RTX A6000\\
    \midrule
    $\alpha_{\kappa}$ &$1.04 \times 10^{-10}$ if $b \leq 8$ & $3.94 \times 10^{-12}$\\
                      &$2.07 \times 10^{-10}$ if $b > 8$  &\\
    $\beta_{\kappa}$  &$3.74 \times 10^{-11}$ if $b \leq 8$ & $1.72 \times 10^{-11}$\\
                      &$-1.60 \times 10^{-9}$ if $b > 8$ &\\
    $\alpha_{\tau}$   & 0                          & $2.07 \times 10^{-13}$\\
    $\beta_{\tau}$    & 0                          & $1.69 \times 10^{-13}$\\
    \bottomrule
  \end{tabularx}
\end{table}

\section{Evaluation}
\label{sec:Evaluation}

\subsection{Evaluation Settings}

The evaluation is conducted on a server equipped with a 36-core Intel(R) Core(TM) i9-10980XE CPU and 128\,GB RAM.

\subsubsection{Model Settings}
\label{sec:Model Settings}

ResNet101~\cite{heDeepResidualLearning2016} is adopted as the global model $\bm{F}$.
Table~\ref{table:model_specification} presents the specifications of the ResNet101 model.
To balance the modularity and granularity of sub-models, each building block is treated as a layer, resulting in $L=37$.
The ImageNet dataset~\cite{dengImageNetLargeScaleHierarchical2009} is used, with input and output dimensions of $3 \times 224 \times 224$ and 1,000, respectively.
Multiply-accumulate operations (MACs) and the output dimension of each layer are measured using the PyTorch library~\cite{paszkePyTorchImperativeStyle2019}.
Forward-computation FLOPs are calculated as twice the MACs, while backward-computation FLOPs are determined as twice the forward-computation FLOPs.
The sizes of activation and gradient data, collectively referred to as smashed data, are calculated based on the output dimensions of each building block, assuming 32-bit floating-point representation for each element.
The memory size $r_l^\mathrm{mem}$ of each layer $l$ $(l = 1, \ldots, L)$ is calculated based on the number of parameters represented as 32-bit floating-point numbers.
For simplicity, it is assumed that the memory size and storage size of each layer $l$ are identical $(r_l^\mathrm{mem} = r_l^\mathrm{disk})$.

In the ResNet101 model, Table~\ref{table:model_specification} reveals the following key characteristics: (C1) Computational costs are substantially higher in the middle layers (layers 3--35); (C2) The smashed data size monotonically decreases as the layers progress, with the exception of the second layer; and (C3) The layer size tends to increase as the layers progress, except for the second layer and the last two layers.

\subsubsection{Network Settings}
\label{sec:Network Settings}

The NSFNET topology~\cite{millsNSFNETBackboneNetwork1987} shown in Fig.~\ref{fig:nsfnet}, consisting of 14 nodes and 42 directed links, is used as the physical network.
Each physical link has a bandwidth of 1\,Gbps.
The link propagation delays are calculated based on the distance and the speed of light in optical fibers, with values ranging from 1.23\,ms to 14.2\,ms.
Each physical node is equipped with available computing resources, as summarized in Table~\ref{table:computing_node_specification}.
To estimate the parameters $\alpha_{\kappa}$, $\beta_{\kappa}$, $\alpha_{\tau}$, and $\beta_{\tau}$, the computation time of ResNet101 is measured while varying the input batch size $b$ from 1 to 256.
Each measurement is repeated 110 times, with the first 10 trials discarded as warm-up.
Assuming that computation time scales linearly with $b$, the measurements are fitted to an OLS linear model to obtain the parameters.
Because the CPU case exhibits different behavior for $b \leq 8$ and $b > 8$, the parameters $\alpha_{\kappa}$ and $\beta_{\kappa}$ are estimated separately for these two ranges.

Each sub-model $\bm{F}^k$, except the first and last sub-models, has a candidate set $\V^k$ of physical nodes for placement, which is randomly and distinctly selected from the physical nodes, where $|\V^k|=2$.
The source node $s$ is constrained by resource limitations to operate solely with a CPU, categorizing it a CPU node, while all other nodes are equipped with GPUs, classifying them as GPU nodes.
Table~\ref{table:computing_node_specification} summarizes the specifications of the computing nodes.
The available memory and storage capacities of these GPU nodes are set to 2\,GB each, while those of the CPU node are set to 8\,GB.
The first and last sub-models are always placed on the source and destination nodes, respectively.

\subsubsection{Schemes for Evaluation}

First, the proposed ILPs ($\mathsf{P}_\mathrm{IF}$ and $\mathsf{P}_\mathrm{TR}$), solved using the Gurobi solver 12.3~\cite{gurobi}, are employed to analyze the characteristics of the optimal solutions. 
Subsequently, the solution quality and computational efficiency of the proposed BCD algorithm are assessed, where the algorithm is implemented in Python 3.12 using the NetworkX library~\cite{hagbergExploringNetworkStructure2008}.
Finally, to verify the importance of jointly optimizing model splitting, placement, and chaining, two comparison schemes are introduced: Computation-oriented model splitting (COMP-MS) and communication-oriented model splitting (COMM-MS).
Both schemes adopt a two-step ILP approach to solve the model splitting, placement, and chaining problem. 
In the first step, the ILP optimizes model splitting to minimize either computation or communication overhead without considering model placement and chaining. 
In the second step, the ILP corresponds to the problem $\mathsf{P}$ based on the optimized model splitting $\bm{y}^*$ obtained from the first step. 

\subsubsection{Evaluation Metrics}

The inference and training latencies per batch are evaluated as the primary metrics.
Additionally, the execution time is measured to assess scalability, where the execution time is defined as the duration required to solve the model splitting, placement, and chaining problem.
If the execution time exceeds the predefined limit of 1,000\,seconds, the process is terminated, and the result is considered infeasible.
All measurements are conducted 10 times, and the average values are reported.

\subsection{Fundamental Characteristics of Optimal Solutions}
\label{sec:Fundamental Characteristics of Optimal Solutions}

\begin{figure*}[!t]
  \centering
  \begin{minipage}{0.48\columnwidth}
    \centering
    \includegraphics[width=\columnwidth]{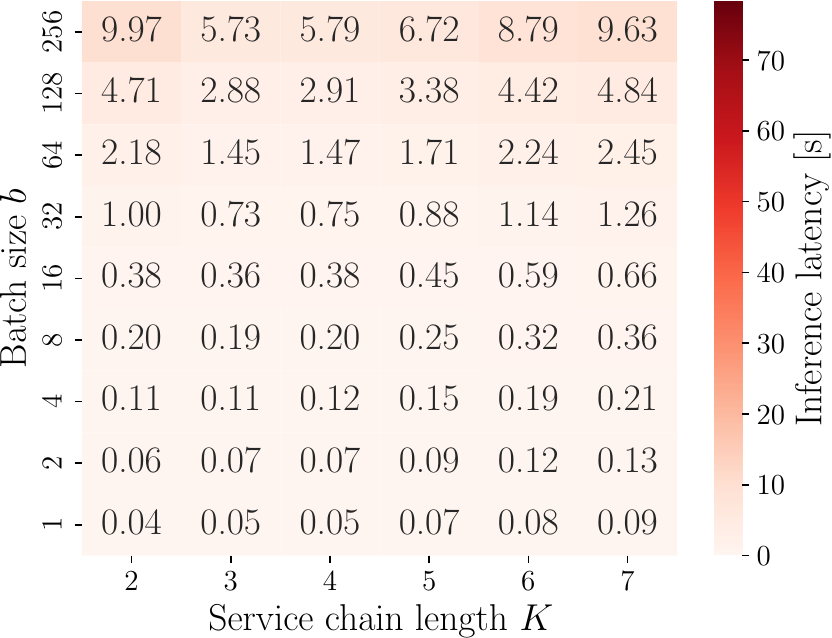}
    \subcaption{ILP.}
    \label{fig:heatmap:msi:ilp}
  \end{minipage}
  \begin{minipage}{0.48\columnwidth}
    \centering
    \includegraphics[width=\columnwidth]{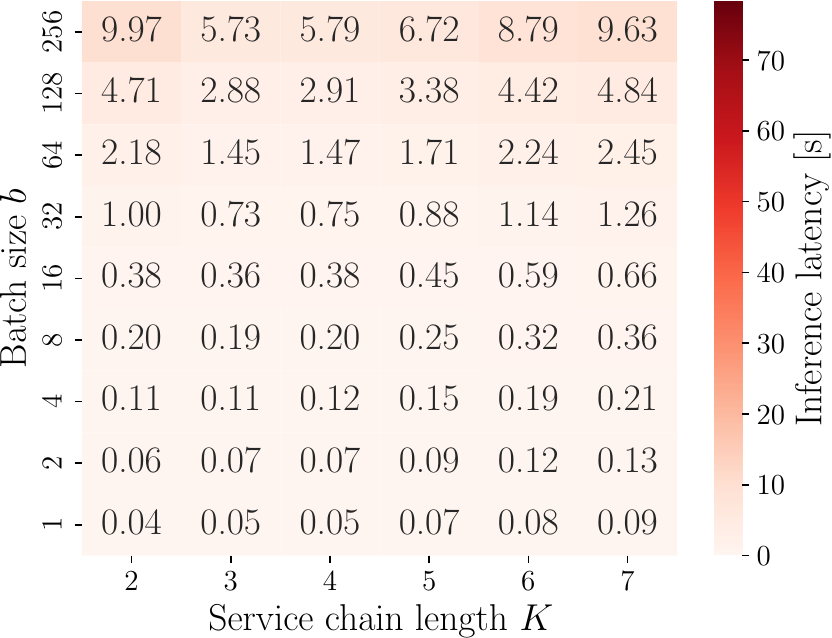}
    \subcaption{BCD.}
    \label{fig:heatmap:msi:bcd}
  \end{minipage}
  \begin{minipage}{0.48\columnwidth}
    \centering
    \includegraphics[width=\columnwidth]{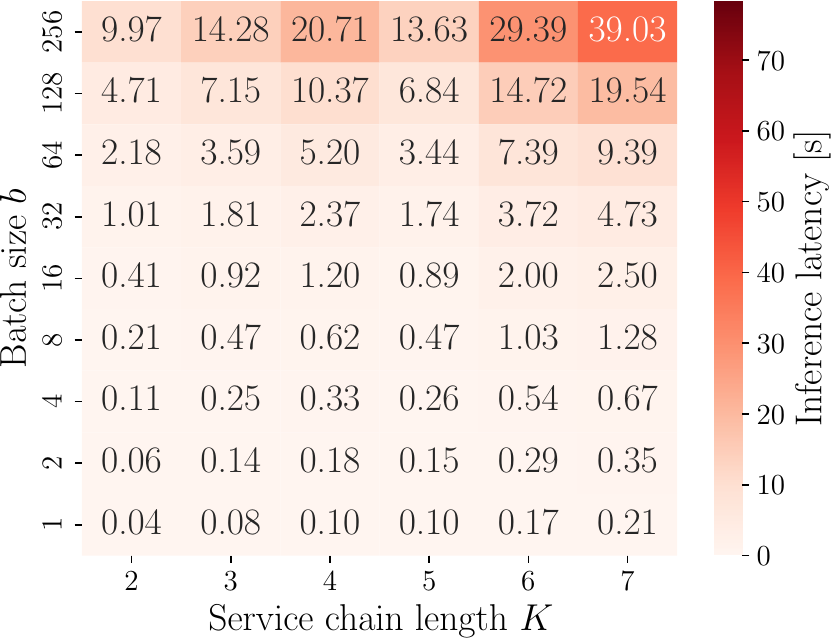}
    \subcaption{COMP-MS.}
    \label{fig:heatmap:msi:comp}
  \end{minipage}
  \begin{minipage}{0.48\columnwidth}
    \centering
    \includegraphics[width=\columnwidth]{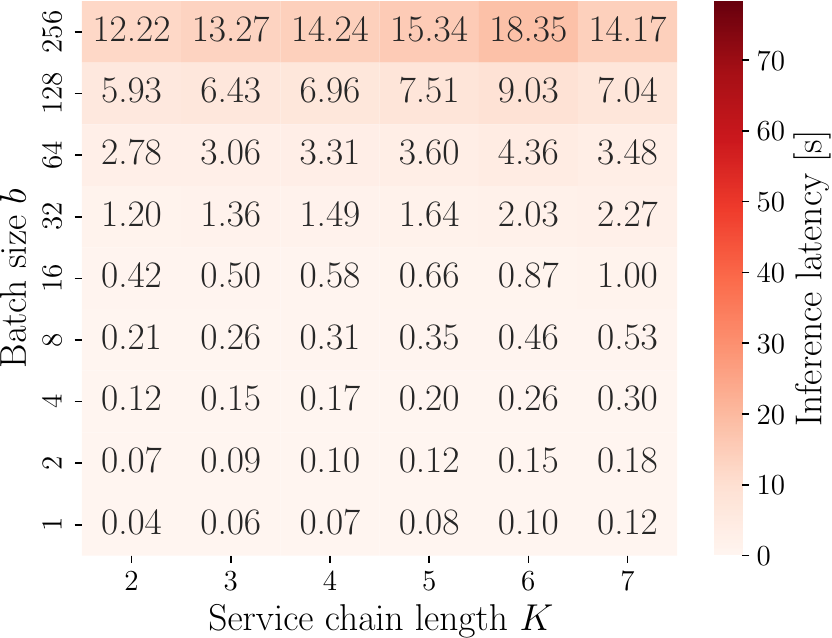}
    \subcaption{COMM-MS.}
    \label{fig:heatmap:msi:comm}
  \end{minipage}
  \caption{Inference latency per batch.}
  \label{fig:heatmap:msi}
\end{figure*}

\begin{figure*}[!t]
  \centering
  \begin{minipage}{0.48\columnwidth}
    \centering
    \includegraphics[width=\columnwidth]{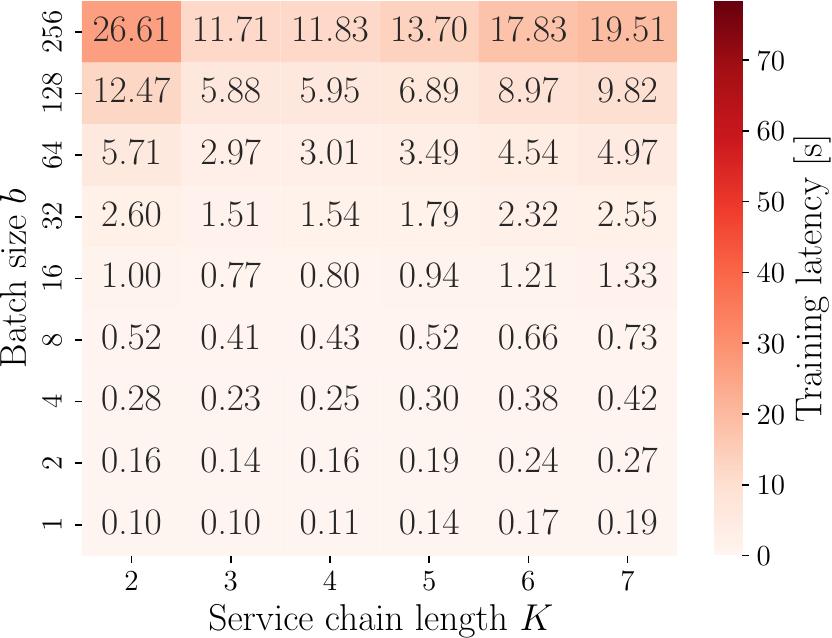}
    \subcaption{ILP.}
    \label{fig:heatmap:msl:ilp}
  \end{minipage}
  \begin{minipage}{0.48\columnwidth}
    \centering
    \includegraphics[width=\columnwidth]{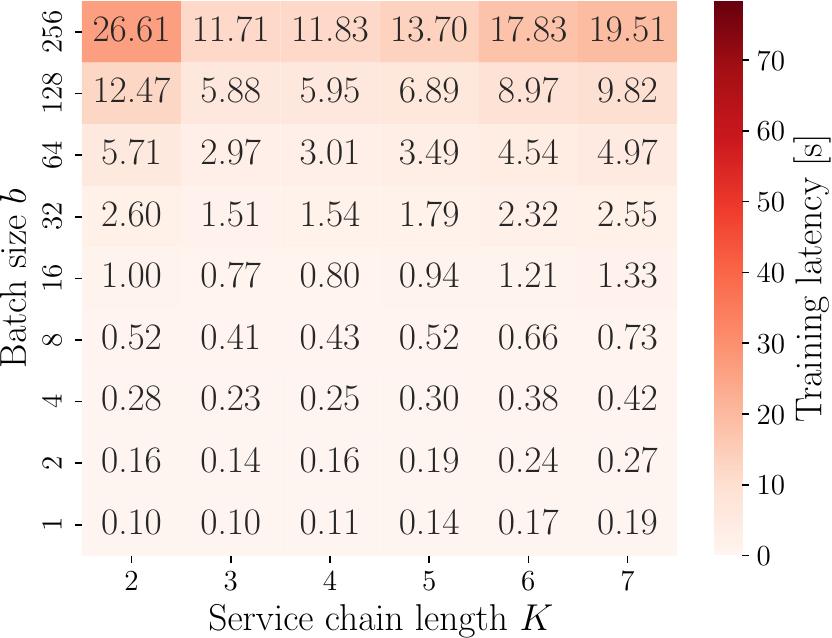}
    \subcaption{BCD.}
    \label{fig:heatmap:msl:bcd}
  \end{minipage}
  \begin{minipage}{0.48\columnwidth}
    \centering
    \includegraphics[width=\columnwidth]{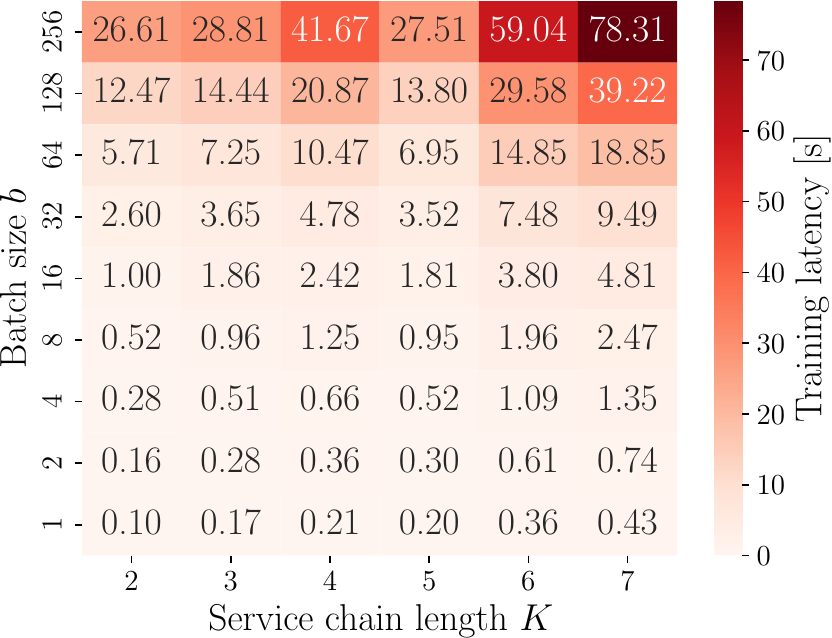}
    \subcaption{COMP-MS.}
    \label{fig:heatmap:msl:comp}
  \end{minipage}
  \begin{minipage}{0.48\columnwidth}
    \centering
    \includegraphics[width=\columnwidth]{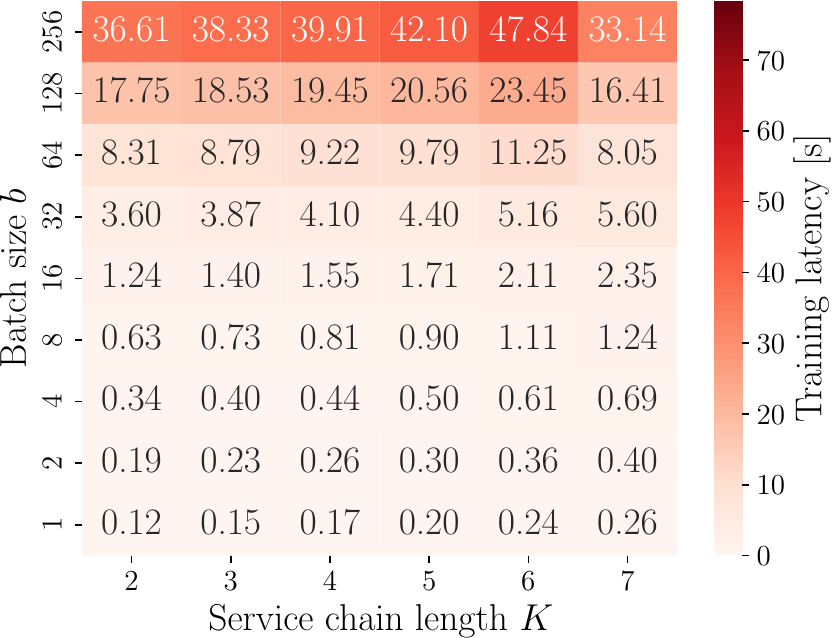}
    \subcaption{COMM-MS.}
    \label{fig:heatmap:msl:comm}
  \end{minipage}
  \caption{Training latency per batch.}
  \label{fig:heatmap:msl}
\end{figure*}

\begin{figure*}[!t]
  \centering
  \begin{minipage}{0.48\columnwidth}
    \centering
    \includegraphics[width=\columnwidth]{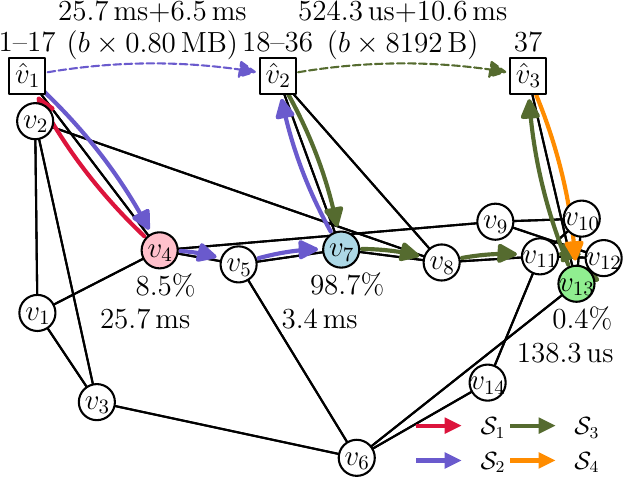}
    \subcaption{ILP.}
    \label{fig:optimal_service_path_and_model_splitting_for_msi:ilp}
  \end{minipage}
  \begin{minipage}{0.48\columnwidth}
    \centering
    \includegraphics[width=\columnwidth]{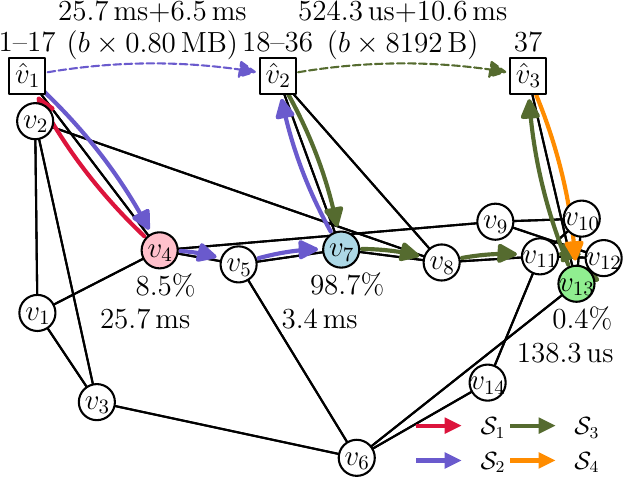}
    \subcaption{BCD.}
    \label{fig:optimal_service_path_and_model_splitting_for_msi:bcd}
  \end{minipage}
  \begin{minipage}{0.48\columnwidth}
    \centering
    \includegraphics[width=\columnwidth]{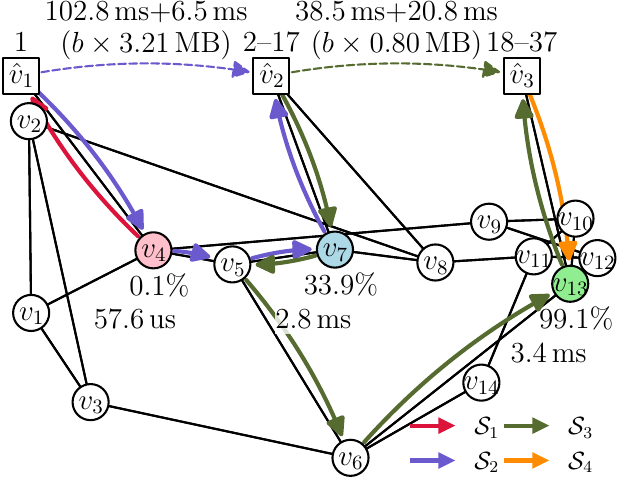}
    \subcaption{COMP-MS.}
    \label{fig:optimal_service_path_and_model_splitting_for_msi:comp}
  \end{minipage}
  \begin{minipage}{0.48\columnwidth}
    \centering
    \includegraphics[width=\columnwidth]{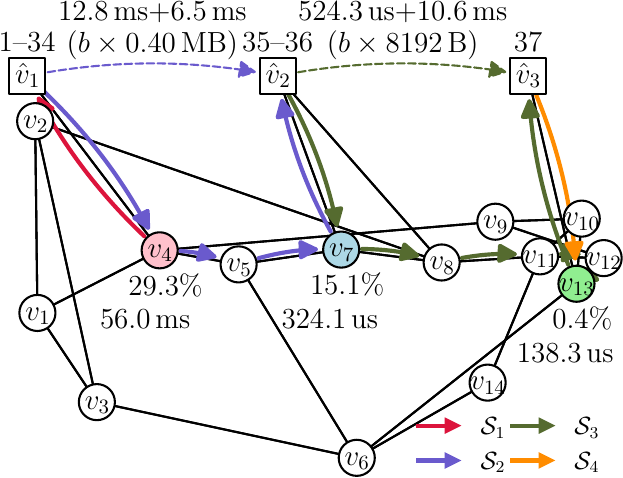}
    \subcaption{COMM-MS.}
    \label{fig:optimal_service_path_and_model_splitting_for_msi:comm}
  \end{minipage}
  \caption{Optimal service path and model splitting for MSI ($K=3$ and $b=2$).}
  \label{fig:optimal_service_path_and_model_splitting_for_msi}
\end{figure*}

\begin{figure*}[!t]
  \centering
  \begin{minipage}{0.48\columnwidth}
    \centering
    \includegraphics[width=\columnwidth]{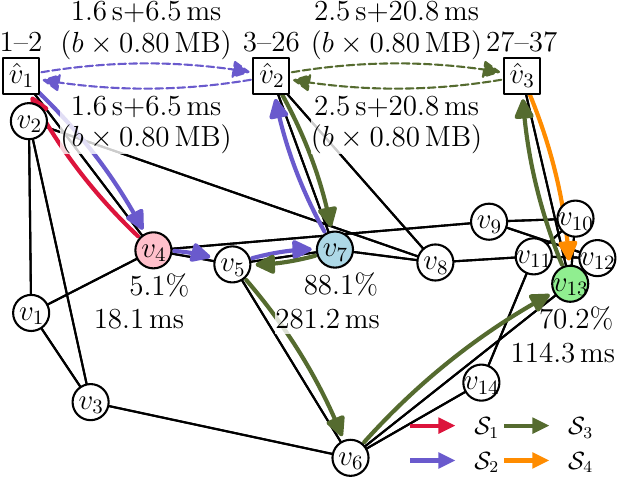}
    \subcaption{ILP.}
    \label{fig:optimal_service_path_and_model_splitting_for_msl:ilp}
  \end{minipage}
  \begin{minipage}{0.48\columnwidth}
    \centering
    \includegraphics[width=\columnwidth]{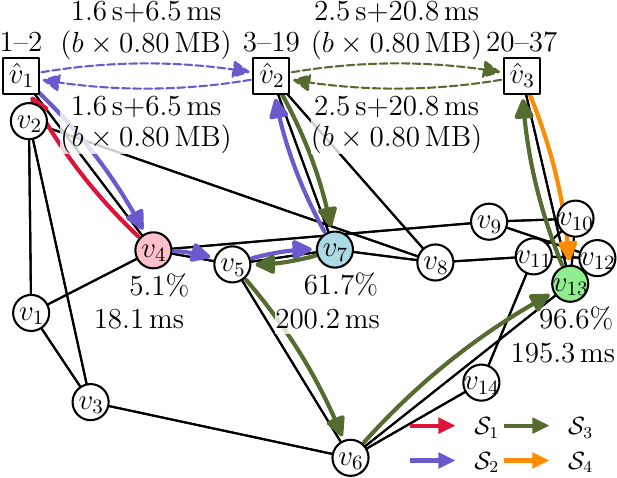}
    \subcaption{BCD.}
    \label{fig:optimal_service_path_and_model_splitting_for_msl:bcd}
  \end{minipage}
  \begin{minipage}{0.48\columnwidth}
    \centering
    \includegraphics[width=\columnwidth]{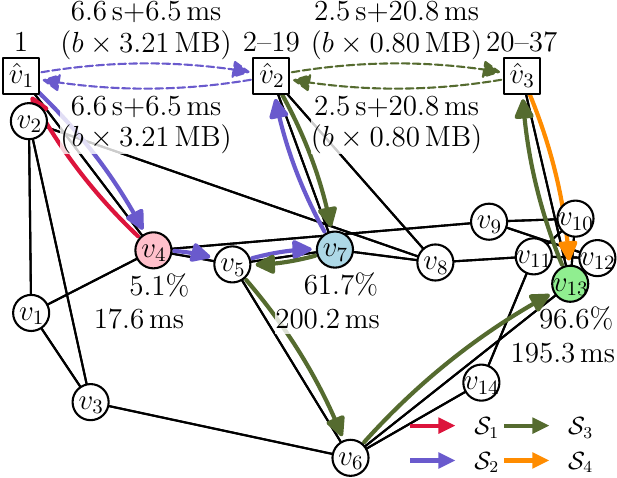}
    \subcaption{COMP-MS.}
    \label{fig:optimal_service_path_and_model_splitting_for_msl:comp}
  \end{minipage}
  \begin{minipage}{0.48\columnwidth}
    \centering
    \includegraphics[width=\columnwidth]{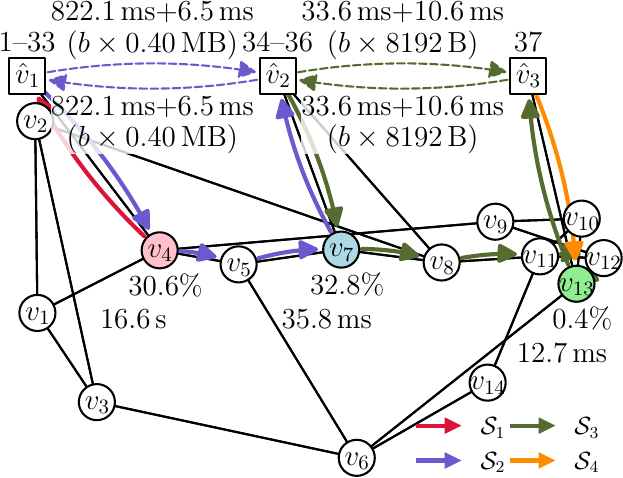}
    \subcaption{COMM-MS.}
    \label{fig:optimal_service_path_and_model_splitting_for_msl:comm}
  \end{minipage}
  \caption{Optimal service path and model splitting for MSL ($K=3$ and $b=128$).}
  \label{fig:optimal_service_path_and_model_splitting_for_msl}
\end{figure*}

\begin{figure}[!t]
  \centering
  \centering
  \includegraphics[width=\columnwidth]{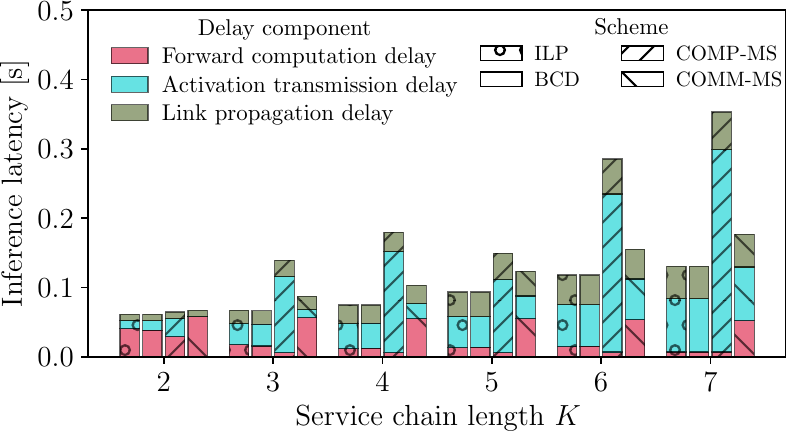}
  \caption{Impact of $K$ on inference latency with breakdown ($b=2$).}
  \label{fig:impact_of_K_on_latency:inference}
\end{figure}
\begin{figure}[!t]
  \centering
  \centering
  \includegraphics[width=\columnwidth]{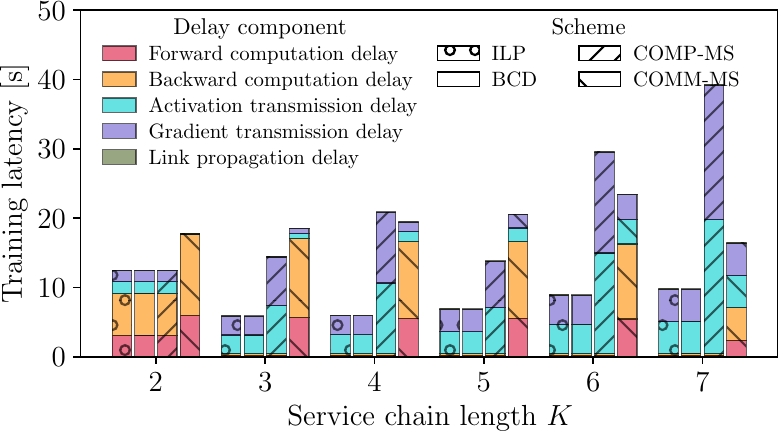}
  \caption{Impact of $K$ on training latency with breakdown ($b=128$).}
  \label{fig:impact_of_K_on_latency:training}
\end{figure}

We first analyze the fundamental characteristics of optimal solutions derived from the ILP.
Figs.~\ref{fig:heatmap:msi:ilp} and \ref{fig:heatmap:msl:ilp} illustrate the inference and training latencies per batch, respectively, for the service chain length $K$ ranging from 2 to 7 in increments of 1, and the batch size $b$ ranging from 1 to 256 in powers of 2.
When focusing on $b$, we observe relatively simple trends.
As $b$ increases, both computation and transmission delays grow, leading to a general increase in inference and training latencies, regardless of $K$.
Additionally, since MSI involves only forwarding processing and communication, while MSL requires bidirectional processing and communication, the inference latency is approximately half of the training latency.

On the other hand, more complex trends are observed with respect to $K$.
For lightweight tasks such as MSI with $b \leq 2$, the optimal configuration is $K=2$, which corresponds to the traditional client-server model splitting approach. 
In contrast, for heavier tasks such as MSI with $b \geq 4$ or MSL, the inference and training latencies are minimized at $K=3$. 
Beyond this point, an increase in $K$ leads to a trend of increasing latency.
Generally, increasing $K$ reduces the computation delay by distributing more computation across GPU nodes.
However, it also increases the communication delay due to the higher number of smashed data transmissions between sub-models.
This not only increases the total propagation delay but also has a more complex effect on the total transmission delay.
Since each physical link has an identical caapcity in this evaluation, the total transmission delay is determined by the cumulative size of the smashed data, which is influenced by the model splitting decisions.

To clarify the impact of computation, transmission, and propagation delays on inference and training latencies, we analyze the impact of $K$ on these latencies with detailed breakdowns for MSI ($b=2$) and MSL ($b=128$), as shown in Figs.~\ref{fig:impact_of_K_on_latency:inference} and \ref{fig:impact_of_K_on_latency:training}, respectively.
For lightweight tasks such as MSI ($b=2$), where the computational load is small, increasing $K$ from 2 to 3 reduces the computation delay by distributing the workload across GPU nodes.
However, this also increases the communication delay due to the additional smashed data transmissions.
As a result, $K=2$ remains slightly more optimal overall, as it minimizes the combined computation and communication overheads.
On the other hand, for heavier tasks such as MSL ($b=128$), where the computational load is significant, it becomes essential to distribute the computational load across GPU nodes by appropriately increasing $K$. 
At the same time, to mitigate the accompanying increase in communication overhead, $K=3$ emerges as the optimal choice. 
Notably, $K=4$ achieves inference and training latencies comparable to $K=3$, but for $K \geq 5$, the impact of increased communication overhead becomes more pronounced.

Model splitting, placement, and chaining decisions are inherently interdependent, as the optimal service path is influenced by the model splitting strategy.
This strategy directly affects computation and communication overheads, as well as memory and storage utilization.
To explore this relationship in detail, we provide examples of the optimal service path and model splitting for MSI and MSL, illustrated in Fig.~\ref{fig:optimal_service_path_and_model_splitting_for_msi:ilp} ($K=3, b=2$) and Fig.~\ref{fig:optimal_service_path_and_model_splitting_for_msl:ilp} ($K=3, b=128$), respectively.

These figures extend Fig.~\ref{fig:nsfnet} by including imaginary nodes (shown as squares), imaginary links (depicted as black lines), the service chain (indicated by colored dashed arrows labeled with transmission delay plus propagation delay and smashed data sizes), subpaths (represented with colored solid arrows), and the allocated model layers displayed above the imaginary nodes.
The source node $v_4$ and destination node $v_{13}$ are highlighted in red and blue, respectively.
Memory utilization and computation delays for executing sub-models are displayed below the respective physical nodes, $v_4$, $v_7$, and $v_{13}$.

In the MSI example of Fig.~\ref{fig:optimal_service_path_and_model_splitting_for_msi:ilp}, the majority of layers are allocated to the source CPU node $v_4$ (i.e., layers 1--17) and the intermediate GPU node $v_7$ (i.e., layers 18--36).
The computational costs for these allocations are 210.60\,GFLOPs and 263.12\,GFLOPs, respectively, with corresponding computation times of 25.7\,ms and 3.4\,ms.
This allocation highlights the advantages of GPU nodes.
While it may seem preferable to allocate more layers to the destination GPU node $v_{13}$, the characteristic (C2) described in Section~\ref{sec:Model Settings} indicates that splitting the model at later layers is more effective in reducing transmission delays caused by the smashed data size.
The comparable computation delay at $v_4$ and the total transmission delay in $\S_{2}$ suggests that the model is split to achieve a balance between computation and transmission delays.
Regarding the subpaths traversing the physical network, $\S_2=(\hat{v}_1, v_4, v_5, v_7, \hat{v}_2)$ represents the minimum-hop path with the smallest propagation delay.
In contrast, $\S_3=(\hat{v}_2, v_7, v_8, v_{11}, v_{12}, v_{13}, \hat{v}_3)$ is not the minimum-hop path but is selected for its minimum propagation delay.
This choice is justified by the extremely small transmission delay per hop, measured at $131.1\,\mathrm{{\textmu}s}$.

In contrast, in the MSL example of  Fig.~\ref{fig:optimal_service_path_and_model_splitting_for_msl:ilp}, the increased task scale leads to a higher computational cost, resulting in a reduced allocation of layers to the source CPU node $v_4$ (i.e., layers 1--2). 
The remaining layers are distributed between the intermediate GPU node $v_7$ (i.e., layers 3--26) and the destination GPU node $v_{13}$ (i.e., layers 27--37) to mitigate the cumulative computation delay. 
Since the GPU nodes have uniform performance in this evaluation, specific model splitting between these nodes does not affect the their cumulative computation delay. 
According to characteristic (C2), allocating more layers to the intermediate GPU node would reduce the smashed data size; however, the memory constraints limit the number of layers assigned to $v_7$. 
Consequently, the computation delays at $v_7$ and $v_{13}$ are 281.2\,ms and 114.3\,ms, respectively, which are significantly higher compared to the MSI example. 
Additionally, the cumulative transmission delays in $\S_2$ and $\S_3$ are 1.6\,s and 2.5\,s, respectively. 
In case of $\S_3$, the minimum-hop path $(\hat{v}_2, v_7, v_5, v_6, v_{13}, \hat{v}_3)$ is selected because, on a per-hop basis, the transmission delay of 833\,ms is significantly dominant compared to the propagation delay of at most 14.2\,ms.

These findings emphasize the necessity of addressing the complex interdependencies among model splitting, placement, and chaining to derive optimal solutions tailored to the specific characteristics and requirements of the given tasks.
\Fin

\subsection{Scheme Comparison}

Next, we assess the effectiveness of the BCD algorithm by comparing its results with the optimal solutions derived from the ILP.
The BCD algorithm achieves almost the same objective values as the ILP solutions for both MSI and MSL, as shown in Figs.~\ref{fig:heatmap:msi:bcd} and \ref{fig:heatmap:msl:bcd}, respectively, by iteratively optimizing the model splitting and service path decisions. 
Furthermore, the BCD algorithm exhibits almost identical trends in the impact of $K$ on these latencies with detailed breakdowns as observed in the ILP solution, as shown in Figs.~\ref{fig:impact_of_K_on_latency:inference} and \ref{fig:impact_of_K_on_latency:training}.
Figs.~\ref{fig:optimal_service_path_and_model_splitting_for_msi:bcd} and \ref{fig:optimal_service_path_and_model_splitting_for_msl:bcd} illustrate the service path and model splitting results of the BCD algorithm for MSI ($K=3, b=2$) and MSL ($K=3, b=128$), respectively. 
Although the splitting points in BCD and ILP for MSL differ, they represent one of the multiple optimal solutions, as layers $11$--$32$ share the same smashed data size of $b \times 0.80\,\mathrm{MB}$.
Similarly, the total computation delay on GPU nodes $v_7$ and $v_{13}$ is identical between the ILP and BCD solutions, indicating an equivalent computational load distributed across these nodes.

To confirm the joint optimization, we further analyze the performance of COMP-MS and COMM-MS relative to the optimal solutions.
Figs.~\ref{fig:heatmap:msi:comp} and \ref{fig:heatmap:msl:comp} illustrate the inference and training latencies of COMP-MS, respectively.
Additionally, the service path and model splitting of COMP-MS are depicted in Fig.~\ref{fig:optimal_service_path_and_model_splitting_for_msi:comp} for MSI ($K=3, b=2$) and Fig.~\ref{fig:optimal_service_path_and_model_splitting_for_msl:comp} for MSL ($K=3, b=128$), respectively.
COMP-MS initially performs model splitting to minimize computation overhead by assigning only the first layer to the source CPU node $v_4$ while distributing the remaining layers across the intermediate and destination GPU nodes, $v_7$ and $v_{13}$.
This results in much smaller computation delays of 57.6\,{\textmu}s and 17.6\,ms on the source CPU node $v_4$ than the ILP solutions for MSI and MSL examples, respectively.
However, this approach also causes a larger smashed data size for $\S_2$ (i.e., $b \times 3.21\,\mathrm{MB}$), reflecting the total transmission delays of $102.8\,\mathrm{ms}$ and $6.6\,\mathrm{s}$ in the subpath $\S_2$ for MSI and MSL examples.
As a consequence, the transmission delay becomes more significant with increasing $K$, and the subsequent step of selecting a service path tends to focus on minimizing the hop count.
This can be confirmed from Figs.~\ref{fig:impact_of_K_on_latency:inference} and \ref{fig:impact_of_K_on_latency:training}, where the transmission delay becomes dominant in the COMP-MS scheme as $K$ increases.
However, we also observe that the overall trend indicates an increase in latency as $K$ grows, with a temporary reduction observed at $K=5$. 
This phenomenon corresponds to changes in the cumulative smashed data size, which directly influences communication overhead. 
Moreover, since COMP-MS exclusively focuses on minimizing computation overhead during model splitting, it may select splitting points that yield identical computation overheads but differ in cumulative smashed data sizes, thereby causing variations in communication overhead.

Figs.~\ref{fig:heatmap:msi:comm} and \ref{fig:heatmap:msl:comm} illustrate the inference and training latencies of COMM-MS, respectively.
The service path and model splitting of COMM-MS are depicted in Fig.~\ref{fig:optimal_service_path_and_model_splitting_for_msi:comm} for MSI ($K=3, b=2$) and Fig.~\ref{fig:optimal_service_path_and_model_splitting_for_msl:comm} for MSL ($K=3, b=128$), respectively.
Since all communication links have uniform capacity in this evaluation, COMM-MS splits the model to minimize the cumulative smashed data size transmitted along subpaths traversing the physical network, specifically $\S_2$ and $\S_3$.
For $K=3$, the model is split such that the smashed data sizes for $\S_2$ and $\S_3$ are $b \times 0.40\,\mathrm{MB}$ and $b \times 8192\,\mathrm{B}$, respectively.
This results in significantly smaller transmission delays of $12.8\,\mathrm{ms}$ and $524.3\,\mathrm{{\textmu}s}$ (resp.\ $822.1\,\mathrm{ms}$ and $33.6\,\mathrm{ms}$) in $\S_2$ and $\S_3$ for the MSI (resp.\ MSL) example, compared to the ILP solutions.
However, this model splitting assigns the first 34 (resp.\ 33) layers for MSI (resp.\ MSL), which constitute the majority of the global model, to the source CPU node $v_4$.
This leads to significantly higher computation delays of 56.0\,ms and 16.6\,s on the $v_4$, as shown in Figs.~\ref{fig:impact_of_K_on_latency:inference} and \ref{fig:impact_of_K_on_latency:training}, respectively.
For $K=7$, even if layers 33--37 are assigned one by one to sub-models after $\hat{v}_3$, it becomes necessary to split layers 1--32 between $\hat{v}_1$ and $\hat{v}_2$. 
Consequently, GPU nodes are utilized more effectively, leading to a reduction in computation delay, particularly in the computationally intensive MSL scenario.

Considering the computational complexity, which will be evaluated in Section~\ref{sec:Scalability}, the BCD algorithm demonstrates a significant advantage by achieving performance comparable to the ILP while drastically reducing the execution time. 
This efficiency is attributed to the decomposition of the original problem into two subproblems, enabling iterative optimization with lower computational overhead. 

\subsection{Scalability Analysis}
\label{sec:Scalability}
\begin{figure}[!t]
  \centering
  \begin{minipage}{0.49\hsize}
    \centering
    \includegraphics[width=\columnwidth]{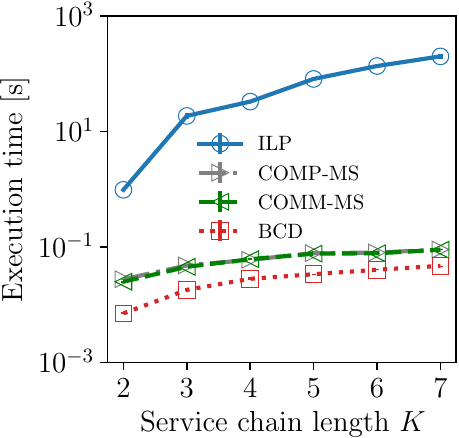}
    \caption{Impact $K$ on execution time.}
    \label{fig:impact_of_K_on_execution_time}
  \end{minipage}
  \begin{minipage}{0.49\hsize}
    \centering
    \includegraphics[width=\columnwidth]{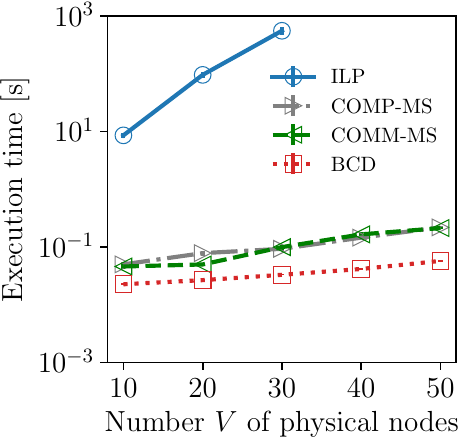}
    \caption{Impact $V$ on execution time.}
    \label{fig:impact_of_V_on_execution_time}
  \end{minipage}
\end{figure}

Fig.~\ref{fig:impact_of_K_on_execution_time} shows the impact of $K$ on the execution time for all schemes.
These results are obtained in the training scenario with $b=128$.
It is noteworthy that the execution time exhibits a similar trend in the inference scenario, irrespective of $b$.
We observe from Fig.~\ref{fig:impact_of_K_on_execution_time} that the execution time increases for all schemes as $K$ grows.
This is attributed to the rise in the number of decision variables and constraints in the ILP formulation with increasing $K$.
Notably, BCD demonstrates a significant advantage, solving the problem over 100 times faster than ILP and consistently finding a feasible solution within 50\,ms.
Furthermore, BCD demonstrates shorter execution times compared to both COMP-MS and COMM-MS.

To evaluate the scalability of the schemes concerning the network size, we consider randomly generated networks with $V$ physical nodes, where each pair of nodes is connected by a physical link with a probability of 0.2.
Fig.~\ref{fig:impact_of_V_on_execution_time} depicts the impact of $V$ on the execution time for all schemes.
The results indicate that ILP fails to find an optimal solution within the predefined execution time limit of 1,000 seconds when $V \geq 30$, whereas the other schemes consistently find feasible solutions.
Additionally, BCD achieves shorter execution times compared to both COMP-MS and COMM-MS, regardless of $V$.
This scalability analysis emphasizes that BCD is more scalable than ILP while maintaining high solution quality.

\section{Conclusion}
\label{sec:Conclusion}

This paper addresses the challenges of model splitting, placement, and chaining in multi-hop split inference and training scenarios.
We formulate the problem as an Integer Linear Programming (ILP) model designed to minimize the training or inference latency.
To improve computational efficiency, we propose a heuristic algorithm based on the block coordinate descent (BCD) method.
This algorithm decomposes the original problem into two subproblems: (1) model splitting and (2) model placement and chaining, and iteratively optimizes these subproblems until convergence. 
Evaluation results reveal the tradeoff between computation and communication across varying mini-batch sizes and service chain lengths.
The findings demonstrate that the proposed BCD achieves inference and training latencies comparable to the optimal ILP solution while significantly reducing computation time. 
Future work will focus on extending the model to address data privacy, multi-path routing, and collaborative model execution across multiple nodes.



\bibliographystyle{IEEEtran}
\bibliography{ref.bib}

@inproceedings{kimBargainingGamePersonalized2023,
  title = {A {{Bargaining Game}} for {{Personalized}}, {{Energy Efficient Split
           Learning}} over {{Wireless Networks}}},
  booktitle = {Proc. of {{IEEE Wireless Communications}} and {{Networking
               Conference}} ({{WCNC}})},
  author = {Kim, Minsu and DeRieux, Alexander and Saad, Walid},
  year = {2023},
  month = mar,
  pages = {1--6},
  issn = {1558-2612},
  doi = {10.1109/WCNC55385.2023.10118601},
  urldate = {2024-11-27},
}

@article{linEfficientParallelSplit2024,
  title = {Efficient {{Parallel Split Learning over Resource-Constrained
           Wireless Edge Networks}}},
  author = {Lin, Zheng and Zhu, Guangyu and Deng, Yiqin and Chen, Xianhao and
            Gao, Yue and Huang, Kaibin and Fang, Yuguang},
  year = {2024},
  month = oct,
  journal = {IEEE Transactions on Mobile Computing},
  volume = {23},
  number = {10},
  pages = {9224--9239},
  issn = {1558-0660},
  doi = {10.1109/TMC.2024.3359040},
  urldate = {2025-02-10},
}

@article{sasabeCapacitatedShortestPath2021,
  title = {Capacitated {Shortest} {Path} {Tour} {Problem}-{Based} {Integer} {
           Linear} {Programming} for {Service} {Chaining} and {Function} {
           Placement} in {NFV} {Networks}},
  volume = {18},
  issn = {1932-4537, 2373-7379},
  doi = {10.1109/TNSM.2020.3044329},
  language = {en},
  number = {1},
  urldate = {2022-10-18},
  journal = {IEEE Transactions on Network and Service Management},
  author = {Sasabe, Masahiro and Hara, Takanori},
  month = mar,
  year = {2021},
  pages = {104--117},
}

@inproceedings{heDeepResidualLearning2016,
  title = {Deep {{Residual Learning}} for {{Image Recognition}}},
  booktitle = {Proc. of {{IEEE Conference}} on {{Computer Vision}} and {{Pattern
               Recognition}} ({{CVPR}})},
  author = {He, Kaiming and Zhang, Xiangyu and Ren, Shaoqing and Sun, Jian},
  year = {2016},
  month = jun,
  pages = {770--778},
  issn = {1063-6919},
  doi = {10.1109/CVPR.2016.90},
  urldate = {2025-02-03},
}

@misc{paszkePyTorchImperativeStyle2019,
  title = {{{PyTorch}}: {{An Imperative Style}}, {{High-Performance Deep
           Learning Library}}},
  shorttitle = {{{PyTorch}}},
  author = {Paszke, Adam and Gross, Sam and Massa, Francisco and Lerer, Adam and
            Bradbury, James and Chanan, Gregory and Killeen, Trevor and Lin,
            Zeming and Gimelshein, Natalia and Antiga, Luca and Desmaison, Alban
            and K{\"o}pf, Andreas and Yang, Edward and DeVito, Zach and Raison,
            Martin and Tejani, Alykhan and Chilamkurthy, Sasank and Steiner,
            Benoit and Fang, Lu and Bai, Junjie and Chintala, Soumith},
  year = {2019},
  month = dec,
  number = {arXiv:1912.01703},
  eprint = {1912.01703},
  primaryclass = {cs, stat},
  publisher = {arXiv},
  doi = {10.48550/arXiv.1912.01703},
  urldate = {2022-10-18},
  archiveprefix = {arXiv},
  langid = {english},
  note = {arXiv:1912.01703},
}

@misc{vepakommaSplitLearningHealth2018,
  title = {Split {{Learning}} for {{Health}}: {{Distributed Deep Learning}}
           without {{Sharing Raw Patient Data}}},
  shorttitle = {Split Learning for Health},
  author = {Vepakomma, Praneeth and Gupta, Otkrist and Swedish, Tristan and
            Raskar, Ramesh},
  year = {2018},
  month = dec,
  number = {arXiv:1812.00564},
  eprint = {1812.00564},
  primaryclass = {cs, stat},
  publisher = {arXiv},
  doi = {10.48550/arXiv.1812.00564},
  urldate = {2024-03-17},
  archiveprefix = {arXiv},
}

@misc{rfc7665,
  series = {Request for Comments},
  number = 7665,
  howpublished = {RFC 7665},
  publisher = {RFC Editor},
  doi = {10.17487/RFC7665},
  url = {https://www.rfc-editor.org/info/rfc7665},
  author = {Joel M. Halpern and Carlos Pignataro},
  title = {{Service Function Chaining (SFC) Architecture}},
  pagetotal = 32,
  year = 2015,
  month = oct,
}

@article{wuSplitLearningWireless2023,
  title = {Split {{Learning over Wireless Networks}}: {{Parallel Design}} and {{
           Resource Management}}},
  shorttitle = {Split {{Learning over Wireless Networks}}},
  author = {Wu, Wen and Li, Mushu and Qu, Kaige and Zhou, Conghao and Shen,
            Xuemin and Zhuang, Weihua and Li, Xu and Shi, Weisen},
  year = {2023},
  month = apr,
  journal = {IEEE Journal on Selected Areas in Communications},
  volume = {41},
  number = {4},
  pages = {1051--1066},
  issn = {1558-0008},
  doi = {10.1109/JSAC.2023.3242704},
  urldate = {2024-11-22},
}

@article{yanOptimalModelPlacement2022,
  title = {Optimal {{Model Placement}} and {{Online Model Splitting}} for {{
           Device-Edge Co-Inference}}},
  author = {Yan, Jia and Bi, Suzhi and Zhang, Ying-Jun Angela},
  year = {2022},
  month = oct,
  journal = {IEEE Transactions on Wireless Communications},
  volume = {21},
  number = {10},
  pages = {8354--8367},
  issn = {1558-2248},
  doi = {10.1109/TWC.2022.3165824},
  urldate = {2025-06-25},
}

@article{zhuESFLEfficientSplit2024,
  title = {{{ESFL}}: {{Efficient Split Federated Learning over
           Resource-Constrained Heterogeneous Wireless Devices}}},
  shorttitle = {{{ESFL}}},
  author = {Zhu, Guangyu and Deng, Yiqin and Chen, Xianhao and Zhang, Haixia and
            Fang, Yuguang and Wong, Tan F.},
  year = {2024},
  month = aug,
  journal = {IEEE Internet of Things Journal},
  volume = {11},
  number = {16},
  pages = {27153--27166},
  issn = {2327-4662},
  doi = {10.1109/JIOT.2024.3397677},
  urldate = {2025-06-25},
}

@misc{gurobi,
  author = {{Gurobi Optimization, LLC}},
  title = {{Gurobi Optimizer Reference Manual}},
  year = 2024,
  url = {https://www.gurobi.com},
}

@inproceedings{millsNSFNETBackboneNetwork1987,
  title = {The {{NSFNET Backbone Network}}},
  booktitle = {Proc. of the {{ACM}} Workshop on {{Frontiers}} in {{Computer
               Communications Technology}}},
  author = {Mills, D. L. and Braun, H.},
  year = {1987},
  month = aug,
  pages = {191--196},
  doi = {10.1145/55482.55502},
  urldate = {2023-07-01},
  isbn = {978-0-89791-245-7},
}

@inproceedings{bhatServiceConcatenationRoutingApplications2017,
  title = {Service-{{Concatenation Routing}} with {{Applications}} to {{Network
           Functions Virtualization}}},
  booktitle = {Proc. of the {{International Conference}} on {{Computer
               Communication}} and {{Networks}} ({{ICCCN}})},
  author = {Bhat, Shireesh and Rouskas, George N.},
  year = {2017},
  month = jul,
  pages = {1--9},
  publisher = {IEEE},
  address = {Vancouver, BC, Canada},
  doi = {10.1109/ICCCN.2017.8038463},
  urldate = {2022-10-18},
  isbn = {978-1-5090-2991-4},
  langid = {english},
}

@article{haraSpeedyEfficientService2022,
  title = {Speedy and {{Efficient Service Chaining}} and {{Function Placement
           Based}} on {{Lagrangian Heuristics}} for {{Capacitated Shortest Path
           Tour Problem}}},
  author = {Hara, Takanori and Sasabe, Masahiro},
  year = {2022},
  month = dec,
  journal = {Journal of Network and Systems Management},
  volume = {31},
  number = {1},
  pages = {24},
  issn = {1573-7705},
  doi = {10.1007/s10922-022-09715-y},
  urldate = {2024-04-14},
  langid = {english},
}

@article{bellmanApproximationCurvesLine1961,
  title = {On the {{Approximation}} of {{Curves}} by {{Line Segments Using
           Dynamic Programming}}},
  author = {Bellman, Richard},
  year = {1961},
  month = jun,
  journal = {Commun. ACM},
  volume = {4},
  number = {6},
  pages = {284},
  issn = {0001-0782},
  doi = {10.1145/366573.366611},
  urldate = {2025-09-07},
}

@article{wrightCoordinateDescentAlgorithms2015,
  title = {Coordinate {{Descent Algorithms}}},
  author = {Wright, Stephen J.},
  year = {2015},
  month = jun,
  journal = {Mathematical Programming},
  volume = {151},
  number = {1},
  pages = {3--34},
  issn = {1436-4646},
  doi = {10.1007/s10107-015-0892-3},
  urldate = {2025-09-07},
  langid = {english},
}

@article{grippoConvergenceBlockNonlinear2000,
  title = {On the {{Convergence}} of the {{Block Nonlinear Gauss}}--{{Seidel
           Method}} under {{Convex Constraints}}},
  author = {Grippo, L. and Sciandrone, M.},
  year = {2000},
  month = apr,
  journal = {Operations Research Letters},
  volume = {26},
  number = {3},
  pages = {127--136},
  issn = {0167-6377},
  doi = {10.1016/S0167-6377(99)00074-7},
  urldate = {2025-09-09},
}

@techreport{hagbergExploringNetworkStructure2008,
  title = {Exploring {{Network Structure}}, {{Dynamics}}, and {{Function Using
           Networkx}}},
  author = {Hagberg, Aric and Swart, Pieter and S Chult, Daniel},
  year = {2008},
  month = jan,
  number = {LA-UR-08-05495; LA-UR-08-5495},
  institution = {Los Alamos National Lab. (LANL), Los Alamos, NM (United States)
                 },
  urldate = {2022-10-18},
  langid = {english},
}

@misc{haraServiceFunctionChaining2025,
  title = {Service {{Function Chaining Architecture}} for {{Multi-hop Split
           Inference}} and {{Learning}}},
  author = {Hara, Takanori and Sasabe, Masahiro},
  year = {2025},
  month = sep,
  number = {arXiv:2509.10001},
  eprint = {2509.10001},
  primaryclass = {cs},
  publisher = {arXiv},
  doi = {10.48550/arXiv.2509.10001},
  urldate = {2025-09-15},
  archiveprefix = {arXiv},
  note = {arXiv:2509.10001},
}

@inproceedings{dengImageNetLargeScaleHierarchical2009,
  title = {{{ImageNet}}: {{A Large-Scale Hierarchical Image Database}}},
  shorttitle = {{{ImageNet}}},
  booktitle = {Proc. of {{IEEE Conference}} on {{Computer Vision}} and {{Pattern
               Recognition}}},
  author = {Deng, Jia and Dong, Wei and Socher, Richard and Li, Li-Jia and Li,
            Kai and {Fei-Fei}, Li},
  year = 2009,
  month = jun,
  pages = {248--255},
  issn = {1063-6919},
  doi = {10.1109/CVPR.2009.5206848},
  urldate = {2026-01-15},
}

@article{jungOptimizationFrameworkSplitting2023,
  title = {{Optimization Framework for Splitting {{DNN}} Inference Jobs over
           Computing Networks}},
  author = {Jung, Sehun and Lee, Hyang-Won},
  year = 2023,
  month = aug,
  journal = {Computer Networks},
  volume = {232},
  pages = {109814},
  issn = {1389-1286},
  doi = {10.1016/j.comnet.2023.109814},
  urldate = {2026-01-29},
}

@inproceedings{liAdaptiveSplitLearning2024,
  title = {Adaptive {{Split Learning}} over {{Energy-Constrained Wireless Edge
           Networks}}},
  booktitle = {{{IEEE INFOCOM}} 2024 - {{IEEE Conference}} on {{Computer
               Communications Workshops}} ({{INFOCOM WKSHPS}})},
  author = {Li, Zuguang and Wu, Wen and Wu, Shaohua and Wang, Wei},
  year = 2024,
  month = may,
  pages = {1--6},
  issn = {2833-0587},
  doi = {10.1109/INFOCOMWKSHPS61880.2024.10620728},
  urldate = {2026-01-29},
}

@article{linHierarchicalSplitFederated2025,
  title = {Hierarchical {{Split Federated Learning}}: {{Convergence Analysis}}
           and {{System Optimization}}},
  shorttitle = {Hierarchical {{Split Federated Learning}}},
  author = {Lin, Zheng and Wei, Wei and Chen, Zhe and Lam, Chan-Tong and Chen,
            Xianhao and Gao, Yue and Luo, Jun},
  year = 2025,
  month = oct,
  journal = {IEEE Transactions on Mobile Computing},
  volume = {24},
  number = {10},
  pages = {9352--9367},
  issn = {1558-0660},
  doi = {10.1109/TMC.2025.3565509},
  urldate = {2026-01-29},
}

@article{marinovaOptimalCutLayer2025,
  title = {Optimal {{Cut Layer Bounds}} for {{Split Learning}}},
  author = {Marinova, Matea and Poposka, Marija and {Hadzi-Velkov}, Zoran and
            Rakovic, Valentin},
  year = 2025,
  month = apr,
  journal = {IEEE Communications Letters},
  volume = {29},
  number = {4},
  pages = {749--753},
  issn = {1558-2558},
  doi = {10.1109/LCOMM.2025.3542541},
  urldate = {2026-01-29},
}

@inproceedings{tiranaEstimatingTrainingTime2025,
  title = {Estimating the {{Training Time}} in {{Single-}} and {{Multi-Hop Split
           Federated Learning}}},
  booktitle = {Proc. of the 8th {{International Workshop}} on {{Edge Systems}},
               {{Analytics}} and {{Networking}}},
  author = {Tirana, Joana and Lalis, Spyros and Chatzopoulos, Dimitris},
  year = 2025,
  month = mar,
  series = {{{EdgeSys}} '25},
  pages = {37--42},
  publisher = {Association for Computing Machinery},
  address = {New York, NY, USA},
  doi = {10.1145/3721888.3722096},
  urldate = {2026-01-29},
  isbn = {979-8-4007-1559-4},
}

@misc{weiPipeliningSplitLearning2025,
  title = {Pipelining {{Split Learning}} in {{Multi-hop Edge Networks}}},
  author = {Wei, Wei and Lin, Zheng and Li, Tao and Li, Xuanheng and Chen,
            Xianhao},
  year = 2025,
  month = sep,
  number = {arXiv:2505.04368},
  eprint = {2505.04368},
  primaryclass = {cs},
  publisher = {arXiv},
  doi = {10.48550/arXiv.2505.04368},
  urldate = {2026-01-29},
  archiveprefix = {arXiv},
}

@article{xuInferenceRoutingMultiHop2026,
  title = {Inference {{Routing over Multi-Hop Edge Networks}}},
  author = {Xu, Ce and Liu, Yuan and Yang, Jiarong},
  year = 2026,
  journal = {IEEE Transactions on Cognitive Communications and Networking},
  volume = {12},
  pages = {1356--1367},
  issn = {2332-7731},
  doi = {10.1109/TCCN.2025.3613490},
  urldate = {2026-01-29},
}

@incollection{vanderbeiLinearProgramming2015,
  title = {{Linear Programming}},
  author = {Vanderbei, Robert J},
  booktitle = {Encyclopedia of Applied and Computational Mathematics},
  pages = {796--800},
  year = {2015},
  publisher = {Springer},
}

@article{tajiriOptimizationDataModel2025,
  title = {Optimization of {{Data}} and {{Model Transfer}} for {{Federated
           Learning}} to {{Manage Large-Scale Network}}},
  author = {Tajiri, Kengo and Kawahara, Ryoichi},
  year = 2025,
  month = apr,
  journal = {IEEE Transactions on Network and Service Management},
  volume = {22},
  number = {2},
  pages = {958--973},
  issn = {1932-4537},
  doi = {10.1109/TNSM.2025.3538156},
  urldate = {2026-01-27},
}

@article{fanDynamicTopologyResource2025,
  title = {Dynamic {{Topology}} and {{Resource Allocation}} for {{Distributed
           Training}} in {{Mobile Edge Computing}}},
  author = {Fan, Weibei and Wang, Donglai and Xiao, Fu and Zuo, Yiping and Lv,
            Mengjie and Han, Lei and Hsieh, Sun-Yuan},
  year = 2025,
  month = jan,
  journal = {IEEE Transactions on Mobile Computing},
  volume = {24},
  number = {11},
  pages = {11927--11941},
  issn = {1558-0660},
  doi = {10.1109/TMC.2025.3581510},
  urldate = {2026-03-17},
}

@article{sartzetakisEdgeCloudInfiniteTime2024,
  title = {Edge/{{Cloud Infinite-Time Horizon Resource Allocation}} for {{
           Distributed Machine Learning}} and {{General Tasks}}},
  author = {Sartzetakis, Ippokratis and Soumplis, Polyzois and Pantazopoulos,
            Panagiotis and Katsaros, Konstantinos V. and Sourlas, Vasilis and
            Varvarigos, Emmanouel},
  year = 2024,
  month = feb,
  journal = {IEEE Transactions on Network and Service Management},
  volume = {21},
  number = {1},
  pages = {697--713},
  issn = {1932-4537},
  doi = {10.1109/TNSM.2023.3312593},
  urldate = {2026-04-16},
}
%

%

\begin{IEEEbiography}[{\includegraphics[width=1in,height=1.25in,clip,keepaspectratio]{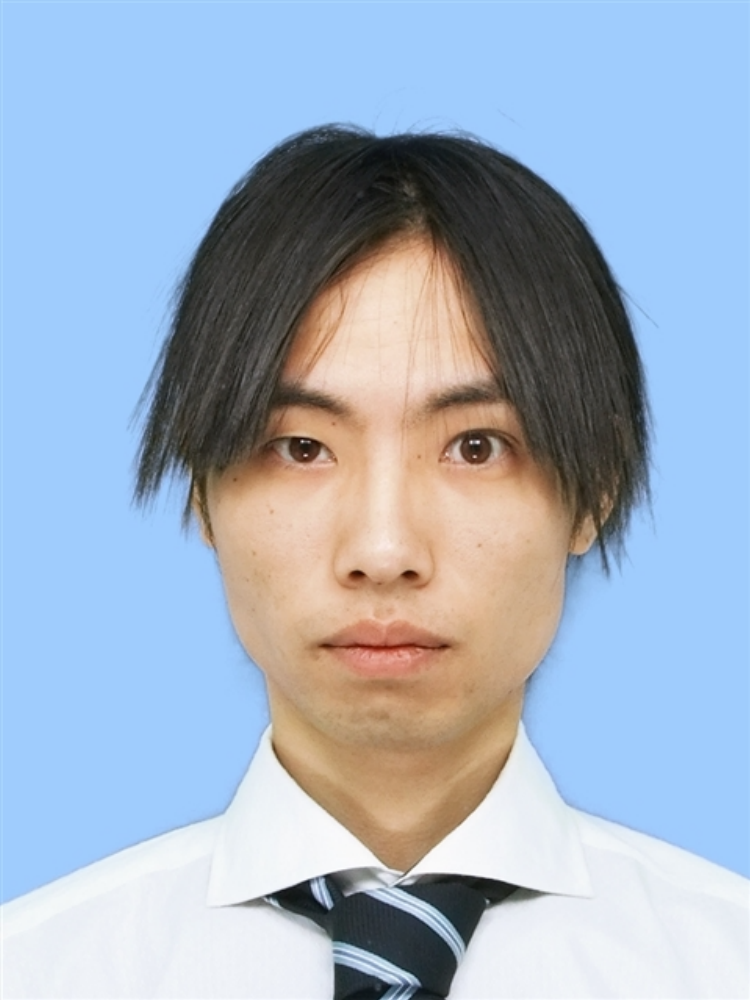}}]{Takanori Hara}
  received the B.Eng. degree from National Institution for Academic Degrees and Quality Enhancement of Higher Education in 2016 and the M.Eng. and Ph.D. degrees from Nara Institute of Science and Technology, Japan, in 2018 and 2021.
  He is currently an Associate Professor with the Division of Information Science, Graduate School of Science and Technology, Nara Institute of Science and Technology, Japan.
  His research interests include eBPF/XDP, NFV, SDN, and networking for AI.
  Dr. Hara is a member of IEEE, ACM, and IEICE.
\end{IEEEbiography}
\begin{IEEEbiography}[{\includegraphics[width=1in,height=1.25in,clip,keepaspectratio]{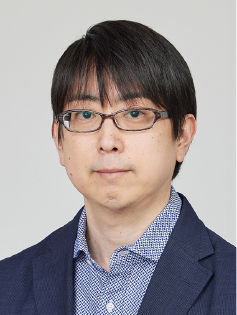}}]{Masahiro Sasabe}
  received the B.S., M.E., and Ph.D. degrees from Osaka University, Japan, in 2001, 2003, and 2006, respectively. He is currently a Professor of Faculty of Informatics, Kansai University, Japan. His research interests include P2P/NFV networking, game-theoretic approaches, human-harmonized network systems, and network optimization.
  Dr. Sasabe is a member of IEEE, ACM, and IEICE.
\end{IEEEbiography}
%

%





\end{document}